\documentclass[aps,prb,preprint,superscriptaddress,showpacs]{revtex4-1}
\usepackage{standalone}
\usepackage{epsfig}
\usepackage{graphicx}
\usepackage{subfigure}
\usepackage{overpic}
\usepackage{array}
\usepackage{xr}
\usepackage{booktabs}
\usepackage{multirow}

\newcommand{\PreserveBackslash}[1]{\let\temp=\\#1\let\\=\temp}
\newcolumntype{C}[1]{>{\PreserveBackslash\centering}p{#1}}
\newcolumntype{R}[1]{>{\PreserveBackslash\raggedleft}p{#1}}
\newcolumntype{L}[1]{>{\PreserveBackslash\raggedright}p{#1}}

\begin{document}
\title{Electronic structures and topological phases of magnetic layered materials MnBi$_2$Te$_4$, MnBi$_2$Se$_4$ and MnSb$_2$Te$_4$}
\author{Ping Li}
\email[]{liping@ahjzu.edu.cn}
\affiliation{Key Laboratory of Advanced Electronic Materials and Devices, School of Physics and Mathematics, Anhui Jianzhu University, Hefei 230601, China}
\author{Jiangying Yu}
\affiliation{Key Laboratory of Advanced Electronic Materials and Devices, School of Physics and Mathematics, Anhui Jianzhu University, Hefei 230601, China}
\author{Ying Wang}
\affiliation{Key Laboratory of Advanced Electronic Materials and Devices, School of Physics and Mathematics, Anhui Jianzhu University, Hefei 230601, China}
\author{Weidong Luo}
\email{wdluo@sjtu.edu.cn}
\affiliation{Key Laboratory of Artificial Structures and Quantum Control, School of Physics and Astronomy, Shanghai Jiao Tong University, Shanghai 200240, China}
\affiliation{Institute of Natural Sciences, Shanghai Jiao Tong University, Shanghai 200240, China}

\date{\today}

\begin{abstract}
First-principles calculations are performed to study the electronic
structures and topological phases of magnetic layered materials
MnBi$_2$Te$_4$, MnBi$_2$Se$_4$ and MnSb$_2$Te$_4$ under different film
thicknesses, strains and spin-orbit coupling (SOC) strengths. All
these compounds energetically prefer the antiferromagnetic (AFM)
state. MnBi$_2$Te$_4$ and MnSb$_2$Te$_4$ bulks are AFM topological
insulators (TIs) in the AFM state, while they become Weyl semimetals
in the ferromagnetic (FM) state. MnBi$_2$Se$_4$ is trivially
insulating in both the AFM and FM states, but it becomes an AFM TI or
a Weyl semimetal with increasing SOC strength or applying compressive
strains. Under equilibrium lattice constants, the FM MnBi$_2$Te$_4$ slabs thicker than two septuple layers
(SLs), the AFM MnBi$_2$Te$_4$ slabs thicker than three SLs and the FM
MnSb$_2$Te$_4$ slabs thicker than five SLs are all Chern
insulators. In addition, Chern insulators can also be obtained by
compressing the in-plane lattice constants of the FM MnBi$_2$Se$_4$
slabs thicker than four SLs and the FM MnSb$_2$Te$_4$ slabs of three
or four SLs, but cannot be obtained using the same method in the AFM
slabs of these two materials. In-plane tensile strains about 1\% to
2\% turn the Chern insulators into trivial insulators.
\end{abstract}

\maketitle
\section{Introduction}
The interplay between topological insulators (TIs) and magnetism is
frequently used to create topological quantum materials such as Chern
insulators, axion insulators and Weyl semimetals \cite{nrp126,
  science895, nature416, nm522, prl206401, sa5685, prl107202,
  prb245209}. These topological phases enable the quantum anomalous
Hall (QAH) effect, the topological magnetoelectric (TME) effect and
chiral Majorana fermions; and they have potential applications in
spintronics and topological quantum computation.  Two methods are
traditionally used to combine magnetism with TIs, i.e., doping
magnetic atoms into TIs \cite{science167, science61, science659} and
constructing TI/magnet heterostructures \cite{nature513, prb085431,
  pb77}. The former tends to introduce inhomogeneity to the doped
bulk, while the latter usually suffers from interfacial band bending
\cite{prb085431}. Hence, the obtained systems are usually far from
satisfactory.

Very recently, MnBi$_2$Te$_4$ has drawn tremendous attention
\cite{science895, nature416, nm522, sa5685,prl206401, prl107202,
  prb195431, sa0948}. This compound is van der Waals (vdW) layered and
intrinsically magnetic. Hence, it naturally avoids both the bulk
inhomogeneity and the interfacial band bending
problems. Three-dimensional (3D) MnBi$_2$Te$_4$ shows rich topological
phases, and in its ground state it is an antiferromagnetic (AFM) TI
\cite{nature416, nm522, sa5685}. The QAH effect has been
experimentally observed in MnBi$_2$Te$_4$ films \cite{science895,
  nm522}, so has the Chern insulator-axion insulator phase transition
\cite{nm522}. Belonging to the same family of materials,
MnBi$_2$Se$_4$ \cite{npjcm33} and MnSb$_2$Te$_4$ \cite{prb195103,
  prb085114} bulks were also investigated and different topological
phases were predicted under various conditions. However, these
predictions are waiting for experimental tests.

The MnBi$_2$Te$_4$ family compounds have small energy gaps
\cite{sa5685, npjcm33, prb085114}. Their band structures and hence the
topological phases strongly depend on the film thicknesses
\cite{prl107202} and applied strains, which must be carefully
considered in experiment. The present work studies the dependence of
MnBi$_2$Te$_4$, MnBi$_2$Se$_4$ and MnSb$_2$Te$_4$ in
detail. First-principles calculations are performed to calculate the
electronic structures and topological phases of these materials under
varying strains and different film thicknesses. The $Z_2$ invariant and the Chern number $C$ are calculated to identify the TIs and Chern insulators, respectively. Special attention is paid to the QAH systems. The chiral edge states and the quantized Hall conductivity $\sigma_{xy}$ of the Chern insulators and the Fermi arcs of the Weyl semimetals are presented. These results are of
significance to the experimental growth and topological-phase control
of the MnBi$_2$Te$_4$ family materials.

\section{Method}
First-principles calculations are performed within the density
functional theory (DFT) framework as implemented in the VASP codes
\cite{prb11169}. The Mn $3d$ orbital is treated by the generalized
gradient approximation \cite{prl3865} plus Hubbard $U$ (GGA+$U$)
method with $U=4.0$ eV. The energy cutoff is 350 eV, and
$14\times14\times2$, $14\times14\times4$ and $9\times9\times1$
Monkhorst-Pack K-meshes are used for the calculations of the AFM
bulks, the FM bulks and the 2D slabs, respectively. The vdW correction
of the optB88-vdW scheme \cite{jpcm022201} is involved to optimize the
crystal structures until the force acting on each ion is less than
0.002 eV/\r{A}. For slabs of different thicknesses, a vacuum of 18
\r{A} is introduced to minimize the image-image interactions due to
the periodic boundary condition. The electronic properties of the AFM
and FM MnBi$_2$Te$_4$ bulks are confirmed by the HSE06 hybrid
functional calculations \cite{jcp8207, jcp219906}, and qualitatively
consistent results are obtained. Topological invariants, surface and
edge states are calculated based on the maximally localized Wannier
functions using the WannierTools package \cite{cpc405}.

\section{Results}
\begin{figure}[htb]
\includegraphics [width=8cm] {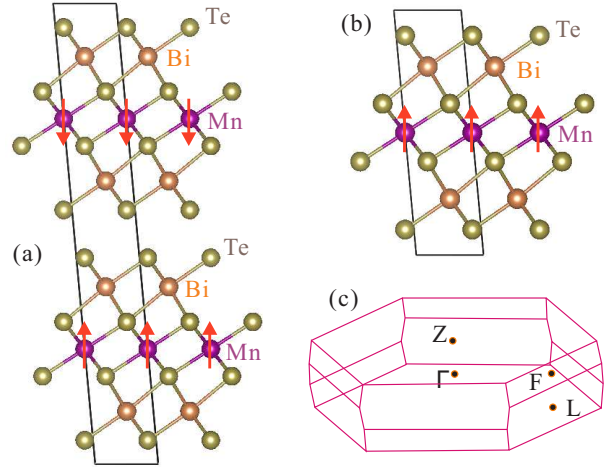}
\caption{Crystal structures of (a) AFM and (b) FM MnBi$_2$Te$_4$, where the arrows indicate the directions of the magnetic moments. (c) Brillouin zone of the crystals.}
\label{crystal}
\end{figure}

MnBi$_2$Te$_4$, MnBi$_2$Se$_4$ and MnSb$_2$Te$_4$ crystallize in the
same rhombohedral structure with the $R\bar{3}m$ space group. We take
MnBi$_2$Te$_4$ as an example and show its structure in
Fig.~\ref{crystal}. It consists of Te-Bi-Te-Mn-Te-Bi-Te septuple
layers (SLs) stacking along the $c$ axis through the vdW interaction.
The calculated lattice constants and the energy differences $\Delta E=E({\rm FM})-E({\rm AFM})$ between the FM and the respective AFM phases are
listed in Table~\ref{table1}, accompanied with previously reported
calculation and experimental data \cite{sa5685, npjcm33,
  prb085114}. The current data agrees well with the previous
reports. The energy differences $\Delta E$ are all positive,
indicating that the AFM state is energetically more favorable, also in
agreement with previous studies \cite{sa5685, npjcm33,prb085114}.

\begin{table*}[htb]
\caption{Lattice constants of MnBi$_2$Te$_4$ family materials and the energy differences $\Delta E=E({\rm FM})-E({\rm AFM})$ between the FM phases and the respective AFM phases. Here, the available data in previous theoretical or experimental reports is presented for comparison. }
\centering  %
\begin{tabular}{L{2.8cm}L{2.2cm}L{3.3cm}L{3.3cm}L{3.3cm}} %
\hline \hline
& &MnBi$_2$Te$_4$ &MnBi$_2$Se$_4$ &MnSb$_2$Te$_4$ \\\hline
\textit{a} (\r{A}) &This work  &4.346   &4.085 &4.281\\
                   &Other calc. &4.36 \cite{sa5685} &4.29, 4.22\cite{npjcm33} &-\\
                   & &4.45, 4.37 \cite{npjcm33} & & \\
                   &Expt. data &4.334 \cite{npjcm33} &4.197 \cite{npjcm33} &4.2385, 4.2219 \cite{prb195103}\\
                   &           &                     &                     &4.2613 \cite{prb085114}\\
\textit{c} (\r{A}) &This work  &41.145  &38.551 &40.510 \\
                   &Other calc. &40.6 \cite{sa5685} &38.94, 38.52 \cite{npjcm33} &-\\
                   & &41.82, 41.38 \cite{npjcm33} & &\\
                   &Expt. data &40.91 \cite{npjcm33} &37.797 \cite{npjcm33} &40.8497, 40.606 \cite{prb195103}\\
                   & & & &41.062 \cite{prb085114}\\
$\Delta E ({\rm meV/Mn})$ &This work &0.3 &0.6 &1.0\\
                   &Other calc. &1.2 \cite{sa5685} &1.5 \cite{npjcm33} &$-$\\\hline
\end{tabular}
\label{table1}
\end{table*}

\subsection{AFM MnBi$_2$Te$_4$, MnBi$_2$Se$_4$ and MnSb$_2$Te$_4$ bulks}
\begin{figure*}[htb]
\includegraphics [width=16cm] {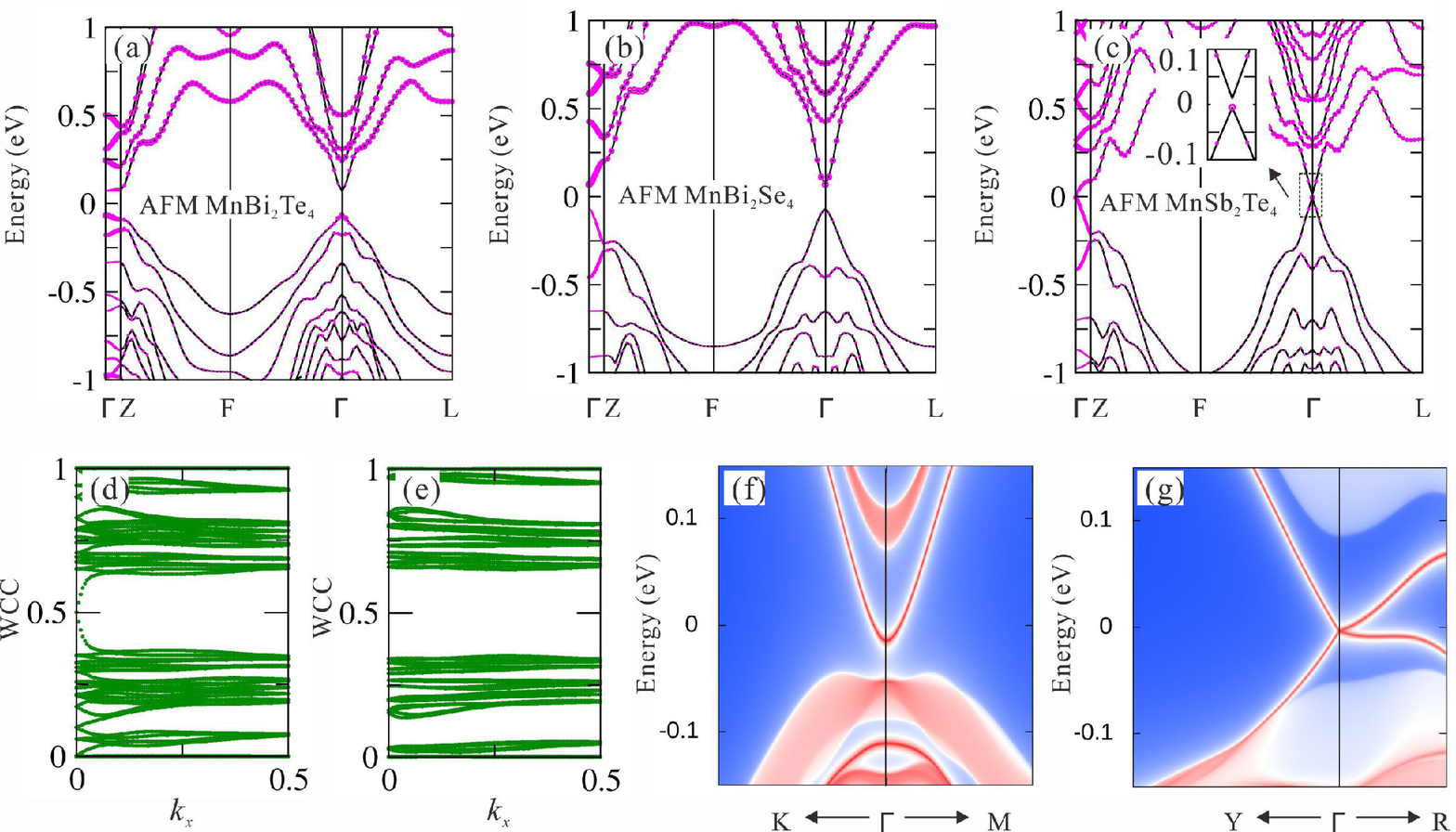}
\caption{Band structures of AFM (a) MnBi$_2$Te$_4$, (b) MnBi$_2$Se$_4$ and (c) MnSb$_2$Te$_4$, where the magenta symbols denote the Bi-$p$ or the Sb-$p$ projections, and the symbol size indicates the contribution weight. Evolution of WCCs of AFM (d) MnBi$_2$Te$_4$ and (e) MnBi$_2$Se$_4$ in the $k_z=0$ plane. Surface states of AFM MnBi$_2$Te$_4$ in (f) (001) and (g) (101) surfaces. }
\label{wcc}
\end{figure*}

Figures~\ref{wcc} (a), (b) and (c) show the calculated band structures of AFM MnBi$_2$Te$_4$, MnBi$_2$Se$_4$ and MnSb$_2$Te$_4$ bulks, respectively. From the figures, the three crystals are all insulating. Both the valence band maxima (VBMs) and the conduction band minima (CBMs) are located at the $\Gamma$ point, forming direct energy gaps of 138, 140 and 16 meV, respectively. Li {\it et al.} \cite{sa5685} predicted energy gaps of about 0.16 and 0.18 eV at the $Z$ and $\Gamma$ points in AFM MnBi$_2$Te$_4$, respectively. $\Gamma$-point direct energy gaps were also theoretically reported in AFM MnBi$_2$Se$_4$ \cite{npjcm33} and MnSb$_2$Te$_4$ \cite{prb085114}. In experiment the measured energy gap of AFM MnBi$_2$Se$_4$ by electrical transport is 0.15 eV \cite{jcg81}, in very good agreement with our data. Figs.~\ref{wcc} (a) and (c) show that the energy bands of AFM MnBi$_2$Te$_4$ and MnSb$_2$Te$_4$ are inverted around the $\Gamma$ point near the Fermi level, while the inversion does not occur in the bands of AFM MnBi$_2$Se$_4$ as Fig.~\ref{wcc} (b) shows. The band-inversion features agree with the previous reports in the MnBi$_2$Te$_4$ case \cite{sa5685} but disagree with the reports in MnBi$_2$Se$_4$ and MnSb$_2$Te$_4$ cases \cite{npjcm33,prb085114}. We attribute the disagreement to the differences of atomic geometries or lattice constants used or obtained in calculations (see Table~\ref{table1}).

Band inversion usually means topologically nontrivial materials, such as 3D TIs \cite{sa5685, prb085114, aqt201900033} and quantum spin Hall (QSH) insulators \cite{sr25423}. The MnBi$_2$Te$_4$ family AFM compounds lose both the time-reversal symmetry ($\Theta$) and the primitive-lattice translational symmetry ($T_{1/2}$) but preserves the combination $S=\Theta T_{1/2}$. Hence, they can be classified according to the the $Z_2$ invariant \cite{prb245209}. To determine their topological nature, the Wannier charge centers (WCCs) are calculated. Figs.~\ref{wcc} (d) and (e) show the evolution of WCCs in the $k_z=0$ plane of MnBi$_2$Te$_4$ and MnBi$_2$Se$_4$, respectively. That of AFM MnSb$_2$Te$_4$ is similar to Fig.~\ref{wcc} (d), thus it is not shown here. Fig.~\ref{wcc} (d) indicates the topological invariant $Z_2=1$ for AFM MnBi$_2$Te$_4$, i.e., it is an AFM TI, so is AFM MnSb$_2$Te$_4$. Fig.~\ref{wcc} (e) gives a topological invariant $Z_2=0$, thus AFM MnBi$_2$Se$_4$ is trivial.
The calculated topological invariants are consistent with the band-inversion features illustrated in Figs.~\ref{wcc} (a) to (c). Different groups have predicted MnBi$_2$Te$_4$ as an AFM TI \cite{sa5685, nature416}, in agreement with the current results. While AFM MnBi$_2$Se$_4$ and MnSb$_2$Te$_4$ bulks were predicted to be topologically nontrivial and trivial \cite{npjcm33,prb085114}, respectively. These differences can be also explained in terms of the differences of lattice constants. According to the calculations by Zhou {\it et al.} \cite{prb085114}, when $c$ is decreased by 6\%, AFM MnSb$_2$Te$_4$ also becomes an AFM TI. The common mechanism of TIs is the spin-orbit coupling (SOC) induced band inversion \cite{rmp3045, rmp1057}. The heavy atoms in MnBi$_2$Te$_4$ produce the strong SOC and a large inverted energy gap of 138 meV. The Sb atom is lighter than the Bi atom, so the SOC is weaker in MnSb$_2$Te$_4$, resulting in an inverted energy gap of only 16 meV. This gap is so small that the topological phase of AFM MnSb$_2$Te$_4$ can be easily changed by external perturbations, such as tensile strains. The SOC of the Se atom is also weaker than that of the Te atom, so that band inversion cannot occur in AFM MnBi$_2$Se$_4$, leading to a trivial state.

In theory, AFM TIs are metallic along some surfaces but insulating along others to support the half-quantum Hall effect with $\sigma_{xy}=e^2/(2h)$ \cite{prb245209}. This is different from strong TIs, which have gapless Dirac surface states along all the surfaces \cite{rmp3045, rmp1057}. Figs.~\ref{wcc} (f) and (g) show the calculated surface states of AFM MnBi$_2$Te$_4$ along the (001) and (101) surfaces, respectively. The former is obviously gapped while the latter exhibits gapless surface states, in consistent with the theory \cite{prb245209}. Moreover, energy distribution curves (EDC) recently resolved an energy gap of about 70 meV at the surface states of the (001) surface of AFM MnBi$_2$Te$_4$ \cite{nature416}, also in agreement with our results.

\begin{figure*}[htb]
\includegraphics [width=16cm] {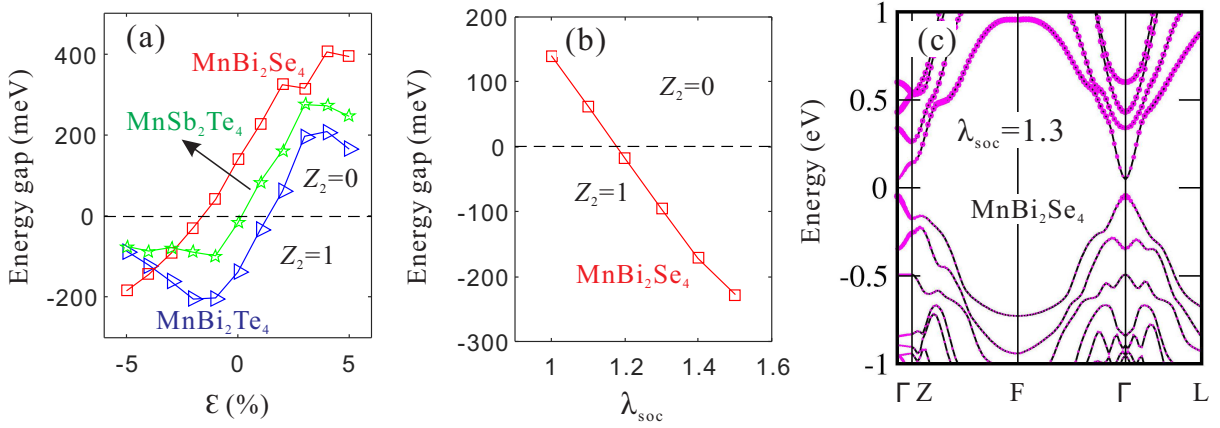}
\caption{(a) Energy gaps and topological phases of AFM MnBi$_2$Te$_4$, MnBi$_2$Se$_4$ and MnSb$_2$Te$_4$ at different strains. (b) Evolution of topological phase and energy gaps of AFM MnBi$_2$Se$_4$ following the SOC strength $\lambda_{\rm soc}$. Positive and negative energy gaps mean that the bulks are trivial insulators with $Z_2=0$ and AFM TIs with $Z_2=1$, respectively, and the absolute values of the gaps correspond to the gap sizes. (c) Band structure of AFM MnBi$_2$Se$_4$ at $\lambda_{\rm soc}=1.3$. }
\label{afmbulkstrain}
\end{figure*}

The electronic structures and consequently the topological phases of materials are closely related to applied strains \cite{aqt201900033, sr25423}. Fig.~\ref{afmbulkstrain} (a) shows the energy gaps and topological phases of AFM MnBi$_2$Te$_4$, MnBi$_2$Se$_4$ and MnSb$_2$Te$_4$ at different strains. Here the isotropic strain is evaluated by
\begin{equation}
\label{eq1}
\begin{array}{l}
 \epsilon=100\times(a-a_0)\%,
\end{array}
\end{equation}
where $a$ and $a_0$ are the strained and unstrained lattice constants, respectively. According to Eq. (\ref{eq1}), a negative $\epsilon$ indicates a compressive strain, while a positive $\epsilon$ means a tensile strain.

Figure~\ref{afmbulkstrain} (a) shows that around the equilibrium lattice constants, the topological phases of the three AFM bulks are very sensitive to applied strains. Under small compressive strains, the energy gaps of AFM MnBi$_2$Te$_4$ and MnSb$_2$Te$_4$ quickly increase but the two bulks remain as AFM TIs. The energy gap of AFM MnBi$_2$Te$_4$ reaches the largest value of 204 meV at $\epsilon=-2\%$. That of AFM MnSb$_2$Te$_4$ rapidly arrives at the maximum of 98 meV at a small compressive strain $\epsilon=-1\%$ from its equilibrium value of 16 meV. Further compressing the bulks, the energy gaps will decrease, but the nontrivial phase can be kept at least for $\epsilon \ge -5\%$. At equilibrium lattice constants, AFM MnBi$_2$Se$_4$ is trivial. From Fig.~\ref{afmbulkstrain} (a), it can be turn into an AFM TI by small or moderate compressive strains. Between $\epsilon=-1\%$ and $-2\%$ the energy gap closes and reopens, and AFM MnBi$_2$Se$_4$ transits from the trivial phase with $Z_2=0$ to the nontrivial phase with $Z_2=1$. It becomes an AFM TI. The nontrivial energy gap at $\epsilon=-2\%$ is 29 meV, which rapidly increases to 184 meV at $\epsilon=-5\%$. On the other hand, tensile strains turn the AFM bulks from the nontrivial phase into the trivial phase. A tensile strain less than 1\% is able to turn AFM MnSb$_2$Te$_4$ to be a trivial insulator because of its rather small equilibrium energy gap (16 meV). At tensile strains around 1\%, AFM MnBi$_2$Te$_4$ keeps in the nontrivial phase. However, at $\epsilon=2\%$ it becomes a trivial insulator with an energy gap of 62 meV.

The SOC strength $\lambda_{\rm soc}$ of AFM MnBi$_2$Se$_4$ is adjusted, and the energy-gap evolution is shown in Fig.~\ref{afmbulkstrain} (b). In this work, $\lambda_{\rm soc}$ is referenced with respect to the original value. For example, $\lambda_{\rm soc}=1.1$ means the SOC strength is adjusted to 1.1 times of the original value. At $\lambda_{\rm soc}=1.0$, the bulk is in a trivial state with $Z_2=0$ and an energy gap of 140 meV. As $\lambda_{\rm soc}$ increases, the energy gap linearly decreases. It closes and reopens near $\lambda_{\rm soc}=1.2$, and phase transition occurs. The compound evolves into a nontrivial state with $Z_2=1$, corresponding to an AFM TI. The phase transition is accompanied by band inversion. Fig.~\ref{afmbulkstrain} (c) shows the band structure of AFM MnBi$_2$Se$_4$ with $\lambda_{\rm soc}=1.3$. Comparing Fig.~\ref{afmbulkstrain} (c) with Fig.~\ref{wcc} (b), band inversion can be clearly resolved around the $\Gamma$ point near the Fermi level.

\subsection{FM MnBi$_2$Te$_4$, MnBi$_2$Se$_4$ and MnSb$_2$Te$_4$ bulks}
The energy differences $\Delta E$ between the AFM and the FM states are very small for the considered bulks (see Table~\ref{table1}), rendering it possible to artificially tune the materials from the AFM state to the FM state. In experiment, such a magnetic phase transition has recently been observed in 6 SLs MnBi$_2$Te$_4$ slabs at a critical magnetic field $\mu_0H_c=4.55$ T \cite{nm522}.

\begin{figure*}[htb]
\includegraphics [width=16cm] {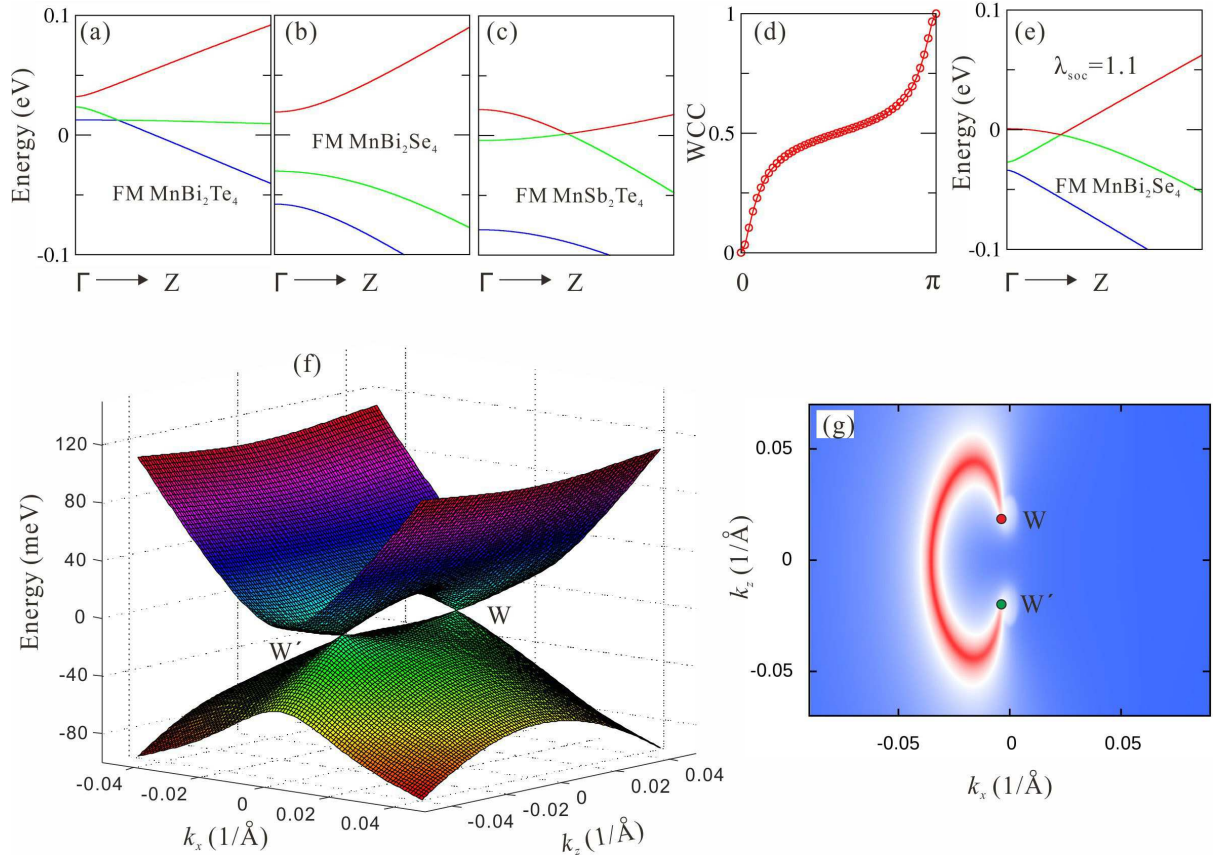}
\caption{Band structures of FM (a) MnBi$_2$Te$_4$, (b) MnBi$_2$Se$_4$ and (c) MnSb$_2$Te$_4$ along the $\Gamma$-Z direction. (d) WCCs Evolution of FM MnSb$_2$Te$_4$. (e) Band structure of FM MnBi$_2$Te$_4$ at $\lambda_{\rm soc}=1.1$. (f) Energy dispersion of FM MnSb$_2$Te$_4$ in the $k_{\rm x}$-$k_{\rm z}$ plane. W and W$'$ indicate the two Weyl points. (g) Fermi arc of FM MnSb$_2$Te$_4$ in the (101) surface. }
\label{arc}
\end{figure*}

Figures~\ref{arc} (a), (b) and (c) show the calculated energy band structures of FM MnBi$_2$Te$_4$, MnBi$_2$Se$_4$ and MnSb$_2$Te$_4$, respectively. From the figures FM MnBi$_2$Te$_4$ and MnSb$_2$Te$_4$ are Weyl semimetals. They each have a pair of Weyl points W and W$'$, which are located on the $-$Z-$\Gamma$-Z line and symmetrical about the $\Gamma$ point. The Weyl points of FM MnBi$_2$Te$_4$ appear at $k_{\rm z}=\pm 0.012$ \r{A}$^{-1}$, and those of FM MnSb$_2$Te$_4$ are at $k_{\rm z}=\pm 0.021$ \r{A}$^{-1}$. To see clearly the Weyl points and the Weyl cones, for instance, we show the energy dispersion of FM MnSb$_2$Te$_4$ in the $k_{\rm x}$-$k_{\rm z}$ plane in Fig.~\ref{arc} (d). According to the WCC calculations, the Weyl chirality of the W point is $+1$, as Fig.~\ref{arc} (e) shows by taking FM MnSb$_2$Te$_4$ as an example, while that of the W$'$ point is $-1$. The present results show that FM MnBi$_2$Te$_4$ is a type-II Weyl semimetal, in agreement with previous calculations \cite{sa5685}, while FM MnSb$_2$Te$_4$ is a type-I Weyl semimetal. We note that previous reports on FM MnSb$_2$Te$_4$ are different. Zhou {\it et al.} \cite{prb085114} predicted an energy gap of 7.75 meV in FM MnSb$_2$Te$_4$, while Murakami {\it et al.} \cite{prb195103} predicted it as a type-II Weyl semimetal. Once again we attribute these differences to the computational details.
The lattice constants $c$ (see Table~\ref{table1}) used by Zhou group \cite{prb085114} and us are 41.062 and 40.510 \r{A}, respectively, compared with the experimental data of 40.8497 and 40.606 \r{A} obtained from XRD and PND techniques \cite{prb195103}, respectively. However, Zhou group \cite{prb085114} reported that FM MnSb$_2$Te$_4$ is a type-I Weyl semimetal under small compressive strains or larger SOC strengths. This is in agreement with our calculations. To clarify whether FM MnSb$_2$Te$_4$ is a Weyl semimetal or not, intensive and delicate experiments are required. Fig.~\ref{arc} (f) demonstrates the calculated Fermi arc on the (101) surface of FM MnSb$_2$Te$_4$, which is the fingerprint of Weyl semimetals.

Different from FM MnBi$_2$Te$_4$ and MnSb$_2$Te$_4$, which are Weyl semimetals, FM MnBi$_2$Se$_4$ is an insulator as Fig.~\ref{arc} (b) shows. However, linear band crossings or Weyl fermions can be obtained by tuning the SOC strength $\lambda_{\rm soc}$ larger. Fig.~\ref{arc} (g) illustrates the band structure of FM MnBi$_2$Se$_4$ along the $\Gamma$-Z line at $\lambda_{\rm soc}=1.1$, where a Weyl point is clearly seen. Increasing $\lambda_{\rm soc}$ from 1.1 to 1.3, the material remains in the Weyl state. Besides, a 1\% compressive strain also turns FM MnBi$_2$Se$_4$ into a type-I Weyl semimetal. The topological phase of FM MnBi$_2$Se$_4$ is different from the calculations of Chowdhury {\it et al.} \cite{npjcm33}, who have predicted it as a Weyl semimetal at equilibrium lattice constants. Moreover, from the figures in that Ref\cite{npjcm33}, it is a type-I Weyl semimetal. This agrees with our calculations. It implies that the difference comes from the differences of the optimized lattice constants. As stated before, topological materials usually have very small energy gaps, thus their topological phases are sensitive to lattice constants or experimental conditions. Currently, the experiments on MnBi$_2$Se$_4$ are quite limited \cite{jcg81}, making the judgement on its topological phase unavailable.

\subsection{MnBi$_2$Te$_4$, MnBi$_2$Se$_4$ and MnSb$_2$Te$_4$ slabs}
The MnBi$_2$Te$_4$ family materials are vdW-layered. They can be mechanically exfoliated into thin flakes from the vdW gaps \cite{nature416, science895, nm522, cpl047301}. The obtained flakes or slabs possess intrinsically magnetic order. According to theoretical studies \cite{prl107202}, the FM exchange interaction between the first nearest neighbors ($J_{01}=0.08$ meV/$\mu_B^2$) strongly dominates over the others in the MnBi$_2$Te$_4$ SL, so the SL exhibits a FM ground state; while the inter-layer coupling is AFM. The intrinsically magnetic order breaks the time-reversal symmetry of the slabs and gaps their surface states, and the half-quantum Hall effect can be expected \cite{prb245209}. The spin-up and spin-down surfaces give rise to Hall conductances $\sigma_{\rm xy}=+e^2/2h$ and $-e^2/2h$, respectively. In the AFM slabs of even SLs, the conductances of different surfaces are cancelled, generating axion insulators. The AFM slabs of odd SLs and FM slabs lead to a total Hall conductance $\sigma_{\rm xy}=e^2/h$, corresponding to QAH insulators with non-zero Chern numbers. The QAH conductance $\sigma_{\rm xy}$ can also be calculated according the Chern number $C$ as \cite{np242}
\begin{equation}
\label{ahcf}
\begin{array}{l}
 \sigma_{\rm xy}=\displaystyle{\frac{e^2}{2\pi h}}\int{dk_{\rm x}dk_{\rm y}b_{\rm z}(\bf k)}=\frac{e^2}{h} C.
\end{array}
\end{equation}
Here $b_{\rm z}({\bf k})$ is the z component the Berry curvature
\begin{equation}
\begin{array}{l}
 b({\bf k})=-{\rm Im}\langle{\partial}_k u({\bf k})|\times|{\partial}_k u({\bf k})\rangle,
\end{array}
\end{equation}
and $u({\bf k})$ is the periodic part of the Bloch function.

\begin{figure*}[hbt]
\includegraphics [width=16cm] {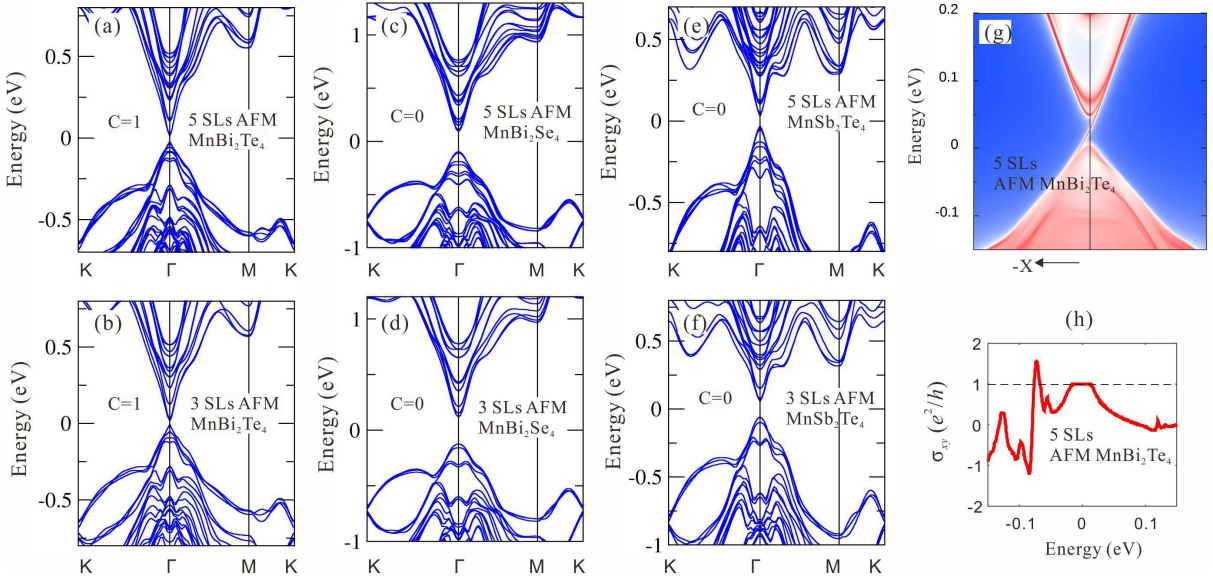}
\caption{(a)-(f) Calculated energy band structures of the AFM MnBi$_2$Te$_4$, MnBi$_2$Se$_4$ and MnSb$_2$Te$_4$ slabs of 3 and 5 SLs, where the Chern numbers are indicated. Calculated (g) chiral edge states and (h) Hall conductance of the AFM MnBi$_2$Te$_4$ slab of 5 SLs.}
\label{afmslabbands}
\end{figure*}
Figure~\ref{afmslabbands} shows the calculated energy band structures of the AFM slabs consisting of 3 and 5 SLs of the three materials, where the Chern numbers are indicated. From Figs.~\ref{afmslabbands} (a) and (b) the AFM MnBi$_2$Te$_4$ slabs are Chern insulators with $C=1$, in agreement with previous reports \cite{prl107202, sa0948}. The energy gaps are 24 and 33 meV at 3 and 5 SLs, respectively, compared with the previously calculated values of 66 and 77 meV \cite{prl107202}, respectively. Larger gaps can be expected for thicker films. According to the picture discussed above, these AFM MnBi$_2$Te$_4$ slabs support chiral edge states and give rise to the QAH effect. The calculated band structure of the AFM MnBi$_2$Te$_4$ ribbon of 5 SLs is illustrated in Fig.~\ref{afmslabbands} (g) as an example, where the chiral edge state is obvious. Tuning the Fermi level into the energy gap, each edge state contributes a Hall conductance $\sigma_{\rm xy}=e^2/h$ in the absence of the external magnetic field. This is the QAH effect, which has potential applications in dissipationless transportation and topological quantum computation \cite{nrp126}. Fig.~\ref{afmslabbands} (h) demonstrates the calculated QAH conductance of the 5 SLs slab. In the energy gap region, the conductance $\sigma_{\rm xy}$ is exactly quantized to $e^2/h$. On the contrary, the AFM MnBi$_2$Te$_4$ slab of 4 SLs has a Chern number $C=0$, and the calculated $\sigma_{\rm xy}$ is zero in the gap region, also in agreement with previous studies \cite{prl107202}. Very recently, the QAH effect has been experimentally observed in the MnBi$_2$Te$_4$ films \cite{science895, nm522}, and the Chern insulator-axion insulator phase transition has also been reported \cite{nm522}. These observations are consistent with our calculations.

Figures~\ref{afmslabbands} (c) to (f) show that the AFM MnBi$_2$Se$_4$ and MnSb$_2$Te$_4$ slabs are trivial insulators with the Chern number $C=0$. This is reasonable. The 3D bulk of AFM MnBi$_2$Se$_4$ is a trivial insulator without band inversion. When exfoliated into AFM slabs band inversion cannot been expected too, so the films remain trivial. The 3D bulk of AFM MnSb$_2$Te$_4$ is an AFM TI with band inversion, but its energy gap is too small, only 16 meV. To realized band inversion in the AFM MnSb$_2$Te$_4$ slabs, the films should be thick enough so that the electronic properties of the films and the bulk converge. Hence, both the AFM MnBi$_2$Se$_4$ and the MnSb$_2$Te$_4$ slabs are not suitable for observing the QAH effect.

\begin{figure*}[htb]
\includegraphics [width=14cm] {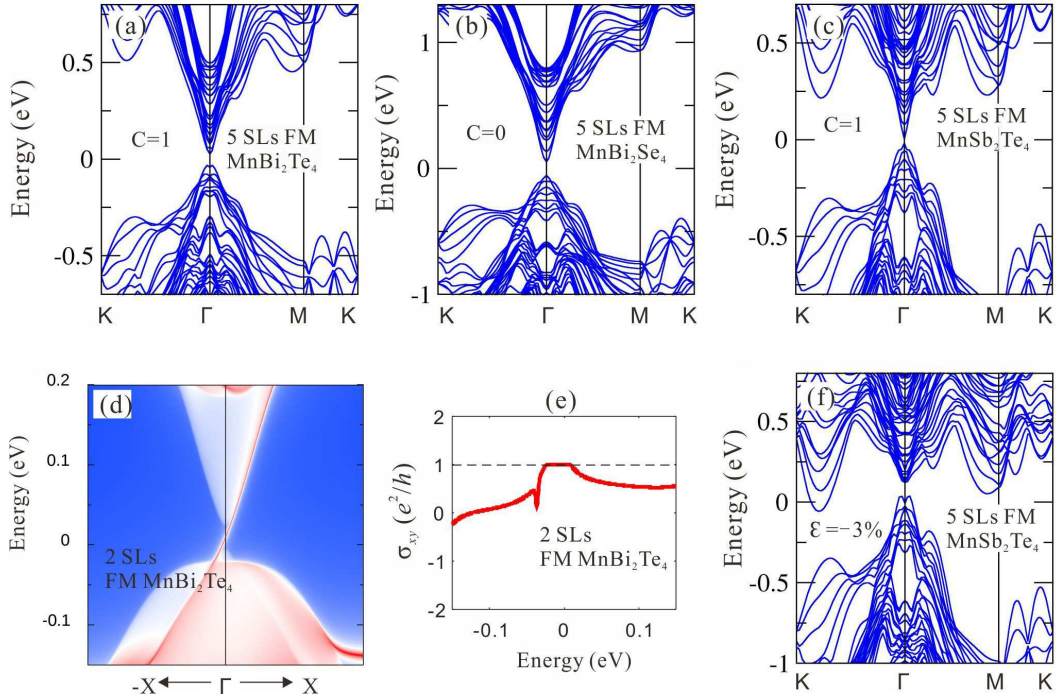}
\caption{Calculated energy band structures of the FM (a) MnBi$_2$Te$_4$, (b) MnBi$_2$Se$_4$ and (c) MnSb$_2$Te$_4$ slabs of 5 SLs, where the Chern numbers are indicated. Calculated (d) chiral edge states and (e) Hall conductance of the FM MnBi$_2$Te$_4$ slab of 2 SLs. (f) Band structure of the FM MnSb$_2$Te$_4$ slab of 5 SLs at $\epsilon=-3\%$. }
\label{fmslabbands}
\end{figure*}
The energy band structures and the topological phases of the FM slabs of these materials are also investigated. At equilibrium lattice constants, the FM slabs have direct energy gaps at the $\Gamma$ point. The FM MnBi$_2$Te$_4$ slabs thicker than 2 SLs and the MnSb$_2$Te$_4$ slabs thicker than 5 SLs behave as Chern insulators with $C=1$, while FM MnBi$_2$Se$_4$ slabs are always trivial insulators with $C=0$ irrespective of the film thickness.
The energy gaps of the FM MnBi$_2$Te$_4$ slabs of 2 to 6 SLs vary from 56 to 72 meV, while the FM MnSb$_2$Te$_4$ slabs of 5 and 6 SLs have gaps of 23 and 41 meV, respectively. The band structures of the FM slabs of 5 SLs are shown in Figs.~\ref{fmslabbands} (a) to (c) as examples. The calculated chiral edge states and the conductivity of the FM MnBi$_2$Te$_4$ slab of 2 SLs are exhibited in Figs.~\ref{fmslabbands} (d) and (e), respectively. The chiral edge state is obvious, and in the energy gap region the QAH conductivity is quantized to $\sigma_{\rm xy}=e^2/h$.
Compared with previous studies, Otrokov {\it et al.} \cite{prl107202} have reported that the FM MnBi$_2$Te$_4$ slab of 2 SLs is a QAH insulator, in agreement with our calculations. Another interesting study theoretically turned the MnBi$_2$Te$_4$ slab of even SLs (4 SLs) from the AFM order to the FM order in a sandwiched CrI$_3$/MnBi$_2$Te$_4$/CrI$_4$ heterostructure, where the QAH effect was predicted \cite{sa0948}.

\begin{figure*}[htb]
\includegraphics [width=16cm] {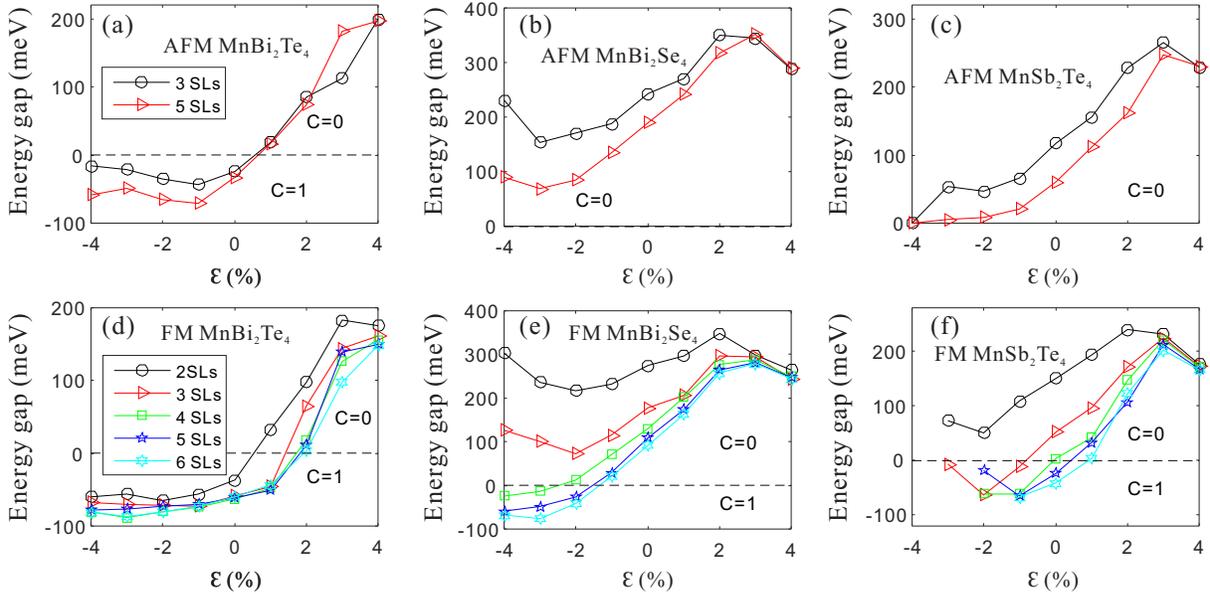}
\caption{Energy gaps and topological phases of (a)-(c) AFM and (d)-(f) FM MnBi$_2$Te$_4$, MnBi$_2$Se$_4$ and MnSb$_2$Te$_4$ slabs at different strains and varying film thicknesses. The legend in (a) applies to (a), (b) and (c), while that in (d) applies to (d), (e) and (f). Positive and negative energy gaps mean that the slabs are trivial insulators with $C=0$ and Chern insulators with $C=1$, respectively, and the absolute values of the gaps correspond to the gap sizes.}
\label{strain}
\end{figure*}
Figure~\ref{strain} shows the energy gaps as well as the topological phases of various AFM and FM slabs at isotropic in-plane strains. All the single layers are trivial insulators with energy gaps of several hundreds of meV. Their behaviours under strains are not shown here because topological phase transitions are hard to be achieved using this method. Thicker slabs behave very differently under in-plane strains. The AFM MnBi$_2$Te$_4$ slabs of 3 and 5 SLs and the FM MnBi$_2$Te$_4$ slabs thicker than 2 SLs are QAH insulators at equilibrium lattice constants with energy gaps ranging from about 20 to 60 meV. Within $-4\% \le \epsilon \le 0$, these slabs keep in the nontrivial state and their energy gaps can be remarkably increased by several tens of meV by in-plane compression. On the other hand, tensile strains from 1\% to 2\% are able to turn the slabs into trivial insulators.

Both the AFM and the FM MnBi$_2$Se$_4$ slabs of 2 SLs or thicker are trivial insulators at $\epsilon =0$. Topological phase transitions from trivial insulators to Chern insulators can be induced in the FM MnBi$_2$Se$_4$ slabs thicker than 4 SLs by applying compressive strains larger than 2\% or 3\%. The AFM and the thinner FM MnBi$_2$Se$_4$ slabs remain trivial when changing $\epsilon$ from $-4\%$ to 4\%. The electronic behaviours of the AFM MnSb$_2$Te$_4$ slabs are similar to those of the AFM MnBi$_2$Se$_4$ slabs. At equilibrium lattice constants, the AFM MnSb$_2$Te$_4$ slabs of 3 and 5 SLs are trivial, and small or moderate strains cannot result in topological phase transitions. The FM MnSb$_2$Te$_4$ slabs of 5 and 6 SLs are Chern insulators with $C=1$, and thinner films are trivial insulators. However, these trivial FM slabs can also be turned into Chern insulators by 1\% to 2\% in-plane compression except the 1 SL slab.
It should be noted that in Fig.~\ref{strain} (f) only the energy gaps and topological phases for $\epsilon \ge -2\%$ or $\epsilon \ge -3\%$ are shown. This is because under larger compressive strains, the CBM becomes lower than the VBM, making the slabs metallic. We show the band structure of the 5 SLs FM MnSb$_2$Te$_4$ slab under $\epsilon=-3\%$ in Fig.~\ref{fmslabbands} (f), which can help us to understand the electronic behaviours of the FM MnSb$_2$Te$_4$ slabs under large compressive strains.

\section{Discussion}

This work provides not only consistent results with previous theoretical studies and experimental observations but also new findings. Both this work and previous studies \cite{nature416, prl206401, sa5685} agree that the MnBi$_2$Te$_4$ bulk is an AFM TI, while in the FM order it is a Weyl semimetal. In experiment, the energy gap has been observed on the (001) surface of MnBi$_2$Te$_4$ \cite{nature416}, and the QAH effect has been realized in a MnBi$_2$Te$_4$ slab of 5 SLs at 1.4 K \cite{science895}. Though the MnBi$_2$Te$_4$ slabs have a AFM ground state, studies on their FM phase is also of interest because their magnetic order is artificially controllable. One experimental example is that the axion insulator to Chern insulator phase transition has been achieved in the MnBi$_2$Te$_4$ slabs of 6 SLs by a moderate magnetic field \cite{nm522}; while an theoretical example is that the MnBi$_2$Te$_4$ slab of even SLs (4 SLs) can be turned from the AFM order to the FM order in a sandwiched CrI$_3$/MnBi$_2$Te$_4$/CrI$_3$ heterostructure \cite{sa0948}. The obtained FM slabs are found to support the QAH effect, also in agreement with our calculations that the FM  MnBi$_2$Te$_4$ slabs thicker than 2 SLs are Chern insulators. In addition, the calculated topological nature of both the AFM and the FM MnBi$_2$Te$_4$ slabs of different SLs also agrees with previous theoretical studies \cite{prl107202, sa0948}. A recent study \cite{nsr1280} has predicted the high-Chern-number Chern insulators in MnBi$_2$Te$_4$ slabs thicker than 8 SLs. The prediction is interesting, and investigations on thicker MnBi$_2$Te$_4$ slabs remain essential.

It seems that the topological phases of the MnSb$_2$Te$_4$ bulk in our work differ from the reports of Zhou {\it et al.} \cite{prb085114}, but in fact they are not contradictory. Zhou {\it et al.} performed calculations based on the experimental lattice constants and the Wyckoff atomic positions, where the Mn and Sb atoms are mixed, while we calculate a perfect MnSb$_2$Te$_4$ crystal at the theoretical lattice constants without any atomic mixing. Therefore, the differences can be explained and understood. Murakami {\it et al.} \cite{prb195103} calculated the properties of MnSb$_2$Te$_4$ and predicted a Weyl semimetal, in agreement with our results. The differences of the topological nature of MnBi$_2$Se$_4$ between our results and the calculations of Chowdhury {\it et al.} \cite{npjcm33} can also be attributed to the atomic-structure differences. Careful experimental studies are needed to clarify the topological properties of MnBi$_2$Se$_4$ and MnSb$_2$Te$_4$.

The topological phases of the considered slabs are sensitive to strains, but as far as we are aware such investigations are currently unavailable. According to our calculations, small isotropic in-plane compressive strains remarkably increase the nontrivial energy gaps of Chern insulators, or change some trivial slabs into QAH insulators. On the contrary, small isotropic in-plane tensile strains around 1\% turn the films from the nontrivial phase to the trivial phase. For the realization and observation of the QAH effect, the MnBi$_2$Te$_4$ slabs are ideal, while the MnBi$_2$Se$_4$ and MnSb$_2$Te$_4$ films are not. On the other hand, the in-plane compressive strains are suggested, while the out-of-plane compressive strains are not recommended. With a common positive Poisson's ratio, the compression of the out-of-plane lattice constant $c$ usually leads to the dilation of the in-plane constant $a$. Our calculations have confirmed this, finding that such operations are adverse to obtain QAH insulators in the present materials.

\section{Conclusion}
The electronic structures and topological phases of magnetic layered materials MnBi$_2$Te$_4$, MnBi$_2$Se$_4$ and MnSb$_2$Te$_4$ are systematically investigated. According to first-principles calculations, these compounds energetically prefer the AFM state. MnBi$_2$Te$_4$ and MnSb$_2$Te$_4$ are AFM TIs in the AFM state, which become Weyl semimetals in the FM state. MnBi$_2$Se$_4$ is trivially insulating in both AFM and FM states, but it can be turned into an AFM TI or a Weyl semimetal by increasing the SOC strength or applying compressive strains. Under equilibrium lattice constants, the FM MnBi$_2$Te$_4$ slabs thicker than 2 SLs and the AFM MnBi$_2$Te$_4$ slabs thicker than 3 SLs are Chern insulators. The AFM MnBi$_2$Se$_4$ and the MnSb$_2$Te$_4$ slabs are trivial insulators, and they are hard to be turned into Chern insulators by applied strains. The FM MnSb$_2$Te$_4$ slabs thicker than 5 SLs are Chern insulators, while the FM MnBi$_2$Se$_4$ slabs are trivial. Some trivial FM slabs can be turned into Chern insulators through small or moderate compressive strains, and some Chern insulating slabs can be easily turned into trivial insulators by 1\% to 2\% tensile strains. The topological phases of MnBi$_2$Te$_4$ have been experimentally confirmed, while those of MnBi$_2$Se$_4$ and MnSb$_2$Te$_4$ have not. In certain cases our calculations give different results from previous calculations, thus they call for careful experimental studies on MnBi$_2$Se$_4$ and MnSb$_2$Te$_4$. The electronic behaviours of the bulks and slabs of all three materials under applied strains and varying film thicknesses are summarized in Fig.~\ref{afmbulkstrain} (a) and Fig.~\ref{strain}. The present work provides references for the experimental control of the topological phases of the MnBi$_2$Te$_4$ family materials, especially for the experimental realization of the QAH effect.

\section*{Acknowledgement}
P. Li would like to thank L. Zhou and J. Li for helpful
discussions. This work is supported by the National Natural Science
Foundation of China (Grant Nos. U1632272 and 11521404), the Natural
Science Foundation of Anhui Province (Grant No. 1308085QA05), and the
Research Foundation of Anhui Jianzhu University (Grant Nos. 2017QD19
and 2018QD11).

\bibliography{Refs}

\begin{thebibliography}{33}%
\makeatletter
\providecommand \@ifxundefined [1]{%
 \@ifx{#1\undefined}
}%
\providecommand \@ifnum [1]{%
 \ifnum #1\expandafter \@firstoftwo
 \else \expandafter \@secondoftwo
 \fi
}%
\providecommand \@ifx [1]{%
 \ifx #1\expandafter \@firstoftwo
 \else \expandafter \@secondoftwo
 \fi
}%
\providecommand \natexlab [1]{#1}%
\providecommand \enquote  [1]{``#1''}%
\providecommand \bibnamefont  [1]{#1}%
\providecommand \bibfnamefont [1]{#1}%
\providecommand \citenamefont [1]{#1}%
\providecommand \href@noop [0]{\@secondoftwo}%
\providecommand \href [0]{\begingroup \@sanitize@url \@href}%
\providecommand \@href[1]{\@@startlink{#1}\@@href}%
\providecommand \@@href[1]{\endgroup#1\@@endlink}%
\providecommand \@sanitize@url [0]{\catcode `\\12\catcode `\$12\catcode
  `\&12\catcode `\#12\catcode `\^12\catcode `\_12\catcode `\%12\relax}%
\providecommand \@@startlink[1]{}%
\providecommand \@@endlink[0]{}%
\providecommand \url  [0]{\begingroup\@sanitize@url \@url }%
\providecommand \@url [1]{\endgroup\@href {#1}{\urlprefix }}%
\providecommand \urlprefix  [0]{URL }%
\providecommand \Eprint [0]{\href }%
\providecommand \doibase [0]{http://dx.doi.org/}%
\providecommand \selectlanguage [0]{\@gobble}%
\providecommand \bibinfo  [0]{\@secondoftwo}%
\providecommand \bibfield  [0]{\@secondoftwo}%
\providecommand \translation [1]{[#1]}%
\providecommand \BibitemOpen [0]{}%
\providecommand \bibitemStop [0]{}%
\providecommand \bibitemNoStop [0]{.\EOS\space}%
\providecommand \EOS [0]{\spacefactor3000\relax}%
\providecommand \BibitemShut  [1]{\csname bibitem#1\endcsname}%
\let\auto@bib@innerbib\@empty
\bibitem [{\citenamefont {Tokura}\ \emph {et~al.}(2019)\citenamefont {Tokura},
  \citenamefont {Yasuda},\ and\ \citenamefont {Tsukazaki}}]{nrp126}%
  \BibitemOpen
  \bibfield  {author} {\bibinfo {author} {\bibfnamefont {Y.}~\bibnamefont
  {Tokura}}, \bibinfo {author} {\bibfnamefont {K.}~\bibnamefont {Yasuda}}, \
  and\ \bibinfo {author} {\bibfnamefont {A.}~\bibnamefont {Tsukazaki}},\ }\href
  {\doibase 10.1038/s42254-018-0011-5} {\bibfield  {journal} {\bibinfo
  {journal} {Nat. Rev. Phys.}\ }\textbf {\bibinfo {volume} {1}},\ \bibinfo
  {pages} {126} (\bibinfo {year} {2019})}\BibitemShut {NoStop}%
\bibitem [{\citenamefont {Deng}\ \emph {et~al.}(2020)\citenamefont {Deng},
  \citenamefont {Yu}, \citenamefont {Shi}, \citenamefont {Guo}, \citenamefont
  {Xu}, \citenamefont {Wang}, \citenamefont {Chen},\ and\ \citenamefont
  {Zhang}}]{science895}%
  \BibitemOpen
  \bibfield  {author} {\bibinfo {author} {\bibfnamefont {Y.}~\bibnamefont
  {Deng}}, \bibinfo {author} {\bibfnamefont {Y.}~\bibnamefont {Yu}}, \bibinfo
  {author} {\bibfnamefont {M.~Z.}\ \bibnamefont {Shi}}, \bibinfo {author}
  {\bibfnamefont {Z.}~\bibnamefont {Guo}}, \bibinfo {author} {\bibfnamefont
  {Z.}~\bibnamefont {Xu}}, \bibinfo {author} {\bibfnamefont {J.}~\bibnamefont
  {Wang}}, \bibinfo {author} {\bibfnamefont {X.~H.}\ \bibnamefont {Chen}}, \
  and\ \bibinfo {author} {\bibfnamefont {Y.}~\bibnamefont {Zhang}},\ }\href
  {\doibase 10.1126/science.aax8156} {\bibfield  {journal} {\bibinfo  {journal}
  {Science}\ }\textbf {\bibinfo {volume} {367}},\ \bibinfo {pages} {895}
  (\bibinfo {year} {2020})}\BibitemShut {NoStop}%
\bibitem [{\citenamefont {Otrokov}\ \emph
  {et~al.}(2019{\natexlab{a}})\citenamefont {Otrokov}, \citenamefont
  {Klimovskikh}, \citenamefont {Bentmann}, \citenamefont {Estyunin},
  \citenamefont {Zeugner}, \citenamefont {Aliev}, \citenamefont {Ga$\beta$},
  \citenamefont {Wolter}, \citenamefont {Koroleva}, \citenamefont {Shikin},
  \citenamefont {Blanco-Rey}, \citenamefont {Hoffmann}, \citenamefont
  {Rusinov}, \citenamefont {Vyazovskaya}, \citenamefont {Eremeev},
  \citenamefont {Koroteev}, \citenamefont {Kuznetsov}, \citenamefont {Freyse},
  \citenamefont {Sanchez-Barriga}, \citenamefont {Amiraslanov}, \citenamefont
  {Babanly}, \citenamefont {Mamedov}, \citenamefont {Abdullayev}, \citenamefont
  {Zverev}, \citenamefont {Alfonsov}, \citenamefont {Kataev}, \citenamefont
  {Buchner}, \citenamefont {Schwier}, \citenamefont {Kumar}, \citenamefont
  {Kimura}, \citenamefont {Petaccia}, \citenamefont {Santo}, \citenamefont
  {Vidal}, \citenamefont {Schatz}, \citenamefont {Ki$\beta$ner}, \citenamefont
  {Unzelmann}, \citenamefont {Min}, \citenamefont {Moser}, \citenamefont
  {Peixoto}, \citenamefont {Reinert}, \citenamefont {Ernst}, \citenamefont
  {Echenique}, \citenamefont {Isaeva},\ and\ \citenamefont
  {Cholkov}}]{nature416}%
  \BibitemOpen
  \bibfield  {author} {\bibinfo {author} {\bibfnamefont {M.~M.}\ \bibnamefont
  {Otrokov}}, \bibinfo {author} {\bibfnamefont {I.~I.}\ \bibnamefont
  {Klimovskikh}}, \bibinfo {author} {\bibfnamefont {H.}~\bibnamefont
  {Bentmann}}, \bibinfo {author} {\bibfnamefont {D.}~\bibnamefont {Estyunin}},
  \bibinfo {author} {\bibfnamefont {A.}~\bibnamefont {Zeugner}}, \bibinfo
  {author} {\bibfnamefont {Z.~S.}\ \bibnamefont {Aliev}}, \bibinfo {author}
  {\bibfnamefont {S.}~\bibnamefont {Ga$\beta$}}, \bibinfo {author}
  {\bibfnamefont {A.~U.~B.}\ \bibnamefont {Wolter}}, \bibinfo {author}
  {\bibfnamefont {A.~V.}\ \bibnamefont {Koroleva}}, \bibinfo {author}
  {\bibfnamefont {A.~M.}\ \bibnamefont {Shikin}}, \bibinfo {author}
  {\bibfnamefont {M.}~\bibnamefont {Blanco-Rey}}, \bibinfo {author}
  {\bibfnamefont {M.}~\bibnamefont {Hoffmann}}, \bibinfo {author}
  {\bibfnamefont {I.~P.}\ \bibnamefont {Rusinov}}, \bibinfo {author}
  {\bibfnamefont {A.~Y.}\ \bibnamefont {Vyazovskaya}}, \bibinfo {author}
  {\bibfnamefont {S.~V.}\ \bibnamefont {Eremeev}}, \bibinfo {author}
  {\bibfnamefont {Y.~M.}\ \bibnamefont {Koroteev}}, \bibinfo {author}
  {\bibfnamefont {V.~M.}\ \bibnamefont {Kuznetsov}}, \bibinfo {author}
  {\bibfnamefont {F.}~\bibnamefont {Freyse}}, \bibinfo {author} {\bibfnamefont
  {J.}~\bibnamefont {Sanchez-Barriga}}, \bibinfo {author} {\bibfnamefont
  {I.~R.}\ \bibnamefont {Amiraslanov}}, \bibinfo {author} {\bibfnamefont
  {M.~B.}\ \bibnamefont {Babanly}}, \bibinfo {author} {\bibfnamefont {N.~T.}\
  \bibnamefont {Mamedov}}, \bibinfo {author} {\bibfnamefont {N.~A.}\
  \bibnamefont {Abdullayev}}, \bibinfo {author} {\bibfnamefont {V.~N.}\
  \bibnamefont {Zverev}}, \bibinfo {author} {\bibfnamefont {A.}~\bibnamefont
  {Alfonsov}}, \bibinfo {author} {\bibfnamefont {V.}~\bibnamefont {Kataev}},
  \bibinfo {author} {\bibfnamefont {B.}~\bibnamefont {Buchner}}, \bibinfo
  {author} {\bibfnamefont {E.~F.}\ \bibnamefont {Schwier}}, \bibinfo {author}
  {\bibfnamefont {S.}~\bibnamefont {Kumar}}, \bibinfo {author} {\bibfnamefont
  {A.}~\bibnamefont {Kimura}}, \bibinfo {author} {\bibfnamefont
  {L.}~\bibnamefont {Petaccia}}, \bibinfo {author} {\bibfnamefont {G.~D.}\
  \bibnamefont {Santo}}, \bibinfo {author} {\bibfnamefont {R.~C.}\ \bibnamefont
  {Vidal}}, \bibinfo {author} {\bibfnamefont {S.}~\bibnamefont {Schatz}},
  \bibinfo {author} {\bibfnamefont {K.}~\bibnamefont {Ki$\beta$ner}}, \bibinfo
  {author} {\bibfnamefont {M.}~\bibnamefont {Unzelmann}}, \bibinfo {author}
  {\bibfnamefont {C.~H.}\ \bibnamefont {Min}}, \bibinfo {author} {\bibfnamefont
  {S.}~\bibnamefont {Moser}}, \bibinfo {author} {\bibfnamefont {T.~R.~F.}\
  \bibnamefont {Peixoto}}, \bibinfo {author} {\bibfnamefont {F.}~\bibnamefont
  {Reinert}}, \bibinfo {author} {\bibfnamefont {A.}~\bibnamefont {Ernst}},
  \bibinfo {author} {\bibfnamefont {P.~M.}\ \bibnamefont {Echenique}}, \bibinfo
  {author} {\bibfnamefont {A.}~\bibnamefont {Isaeva}}, \ and\ \bibinfo {author}
  {\bibfnamefont {E.~V.}\ \bibnamefont {Cholkov}},\ }\href {\doibase
  10.1038/s41586-019-1840-9} {\bibfield  {journal} {\bibinfo  {journal}
  {Nature}\ }\textbf {\bibinfo {volume} {576}},\ \bibinfo {pages} {416}
  (\bibinfo {year} {2019}{\natexlab{a}})}\BibitemShut {NoStop}%
\bibitem [{\citenamefont {Liu}\ \emph {et~al.}(2020)\citenamefont {Liu},
  \citenamefont {Wang}, \citenamefont {Li}, \citenamefont {Wu}, \citenamefont
  {Li}, \citenamefont {Li}, \citenamefont {He}, \citenamefont {Xu},
  \citenamefont {Zhang},\ and\ \citenamefont {Wang}}]{nm522}%
  \BibitemOpen
  \bibfield  {author} {\bibinfo {author} {\bibfnamefont {C.}~\bibnamefont
  {Liu}}, \bibinfo {author} {\bibfnamefont {Y.}~\bibnamefont {Wang}}, \bibinfo
  {author} {\bibfnamefont {H.}~\bibnamefont {Li}}, \bibinfo {author}
  {\bibfnamefont {Y.}~\bibnamefont {Wu}}, \bibinfo {author} {\bibfnamefont
  {Y.}~\bibnamefont {Li}}, \bibinfo {author} {\bibfnamefont {J.}~\bibnamefont
  {Li}}, \bibinfo {author} {\bibfnamefont {K.}~\bibnamefont {He}}, \bibinfo
  {author} {\bibfnamefont {Y.}~\bibnamefont {Xu}}, \bibinfo {author}
  {\bibfnamefont {J.}~\bibnamefont {Zhang}}, \ and\ \bibinfo {author}
  {\bibfnamefont {Y.}~\bibnamefont {Wang}},\ }\href {\doibase
  10.1038/s41563-019-0573-3} {\bibfield  {journal} {\bibinfo  {journal} {Nat.
  Mater.}\ }\textbf {\bibinfo {volume} {19}},\ \bibinfo {pages} {522} (\bibinfo
  {year} {2020})}\BibitemShut {NoStop}%
\bibitem [{\citenamefont {Zhang}\ \emph {et~al.}(2019)\citenamefont {Zhang},
  \citenamefont {Shi}, \citenamefont {Zhu}, \citenamefont {Xing}, \citenamefont
  {Zhang},\ and\ \citenamefont {Wang}}]{prl206401}%
  \BibitemOpen
  \bibfield  {author} {\bibinfo {author} {\bibfnamefont {D.}~\bibnamefont
  {Zhang}}, \bibinfo {author} {\bibfnamefont {M.}~\bibnamefont {Shi}}, \bibinfo
  {author} {\bibfnamefont {T.}~\bibnamefont {Zhu}}, \bibinfo {author}
  {\bibfnamefont {D.}~\bibnamefont {Xing}}, \bibinfo {author} {\bibfnamefont
  {H.}~\bibnamefont {Zhang}}, \ and\ \bibinfo {author} {\bibfnamefont
  {J.}~\bibnamefont {Wang}},\ }\href {\doibase 10.1103/PhysRevLett.122.206401}
  {\bibfield  {journal} {\bibinfo  {journal} {Phys. Rev. Lett.}\ }\textbf
  {\bibinfo {volume} {122}},\ \bibinfo {pages} {206401} (\bibinfo {year}
  {2019})}\BibitemShut {NoStop}%
\bibitem [{\citenamefont {Li}\ \emph {et~al.}(2019{\natexlab{a}})\citenamefont
  {Li}, \citenamefont {Li}, \citenamefont {Du}, \citenamefont {Wang},
  \citenamefont {Gu}, \citenamefont {Zhang}, \citenamefont {He}, \citenamefont
  {Duan},\ and\ \citenamefont {Xu}}]{sa5685}%
  \BibitemOpen
  \bibfield  {author} {\bibinfo {author} {\bibfnamefont {J.}~\bibnamefont
  {Li}}, \bibinfo {author} {\bibfnamefont {Y.}~\bibnamefont {Li}}, \bibinfo
  {author} {\bibfnamefont {S.}~\bibnamefont {Du}}, \bibinfo {author}
  {\bibfnamefont {Z.}~\bibnamefont {Wang}}, \bibinfo {author} {\bibfnamefont
  {B.-L.}\ \bibnamefont {Gu}}, \bibinfo {author} {\bibfnamefont {S.-C.}\
  \bibnamefont {Zhang}}, \bibinfo {author} {\bibfnamefont {K.}~\bibnamefont
  {He}}, \bibinfo {author} {\bibfnamefont {W.}~\bibnamefont {Duan}}, \ and\
  \bibinfo {author} {\bibfnamefont {Y.}~\bibnamefont {Xu}},\ }\href {\doibase
  10.1126/sicadv.aaw5685} {\bibfield  {journal} {\bibinfo  {journal} {Sci.
  Adv.}\ }\textbf {\bibinfo {volume} {5}},\ \bibinfo {pages} {eaaw5685}
  (\bibinfo {year} {2019}{\natexlab{a}})}\BibitemShut {NoStop}%
\bibitem [{\citenamefont {Otrokov}\ \emph
  {et~al.}(2019{\natexlab{b}})\citenamefont {Otrokov}, \citenamefont {Rusinov},
  \citenamefont {Blanco-Rey}, \citenamefont {Hoffmann}, \citenamefont
  {Vyazovskaya}, \citenamefont {Eremeev}, \citenamefont {Ernst}, \citenamefont
  {Echenique}, \citenamefont {Arnau},\ and\ \citenamefont
  {Cholkov}}]{prl107202}%
  \BibitemOpen
  \bibfield  {author} {\bibinfo {author} {\bibfnamefont {M.~M.}\ \bibnamefont
  {Otrokov}}, \bibinfo {author} {\bibfnamefont {I.~P.}\ \bibnamefont
  {Rusinov}}, \bibinfo {author} {\bibfnamefont {M.}~\bibnamefont {Blanco-Rey}},
  \bibinfo {author} {\bibfnamefont {M.}~\bibnamefont {Hoffmann}}, \bibinfo
  {author} {\bibfnamefont {A.~Y.}\ \bibnamefont {Vyazovskaya}}, \bibinfo
  {author} {\bibfnamefont {S.~V.}\ \bibnamefont {Eremeev}}, \bibinfo {author}
  {\bibfnamefont {A.}~\bibnamefont {Ernst}}, \bibinfo {author} {\bibfnamefont
  {P.~M.}\ \bibnamefont {Echenique}}, \bibinfo {author} {\bibfnamefont
  {A.}~\bibnamefont {Arnau}}, \ and\ \bibinfo {author} {\bibfnamefont {E.~V.}\
  \bibnamefont {Cholkov}},\ }\href {\doibase 10.1103/PhysRevLett.122.107202}
  {\bibfield  {journal} {\bibinfo  {journal} {Phys. Rev. Lett.}\ }\textbf
  {\bibinfo {volume} {122}},\ \bibinfo {pages} {107020} (\bibinfo {year}
  {2019}{\natexlab{b}})}\BibitemShut {NoStop}%
\bibitem [{\citenamefont {Mong}\ \emph {et~al.}(2010)\citenamefont {Mong},
  \citenamefont {Essin},\ and\ \citenamefont {Moore}}]{prb245209}%
  \BibitemOpen
  \bibfield  {author} {\bibinfo {author} {\bibfnamefont {R.~S.~K.}\
  \bibnamefont {Mong}}, \bibinfo {author} {\bibfnamefont {A.~M.}\ \bibnamefont
  {Essin}}, \ and\ \bibinfo {author} {\bibfnamefont {J.~E.}\ \bibnamefont
  {Moore}},\ }\href {\doibase 10.1103/PhysRevB.81.245209} {\bibfield  {journal}
  {\bibinfo  {journal} {Phys. Rev. B}\ }\textbf {\bibinfo {volume} {81}},\
  \bibinfo {pages} {245209} (\bibinfo {year} {2010})}\BibitemShut {NoStop}%
\bibitem [{\citenamefont {Chang}\ \emph {et~al.}(2013)\citenamefont {Chang},
  \citenamefont {Zhang}, \citenamefont {Feng}, \citenamefont {Shen},
  \citenamefont {Zhang}, \citenamefont {Guo}, \citenamefont {Li}, \citenamefont
  {Ou}, \citenamefont {Wei}, \citenamefont {Wang}, \citenamefont {Ji},
  \citenamefont {Feng}, \citenamefont {Ji}, \citenamefont {Chen}, \citenamefont
  {Jia}, \citenamefont {Dai}, \citenamefont {Fang}, \citenamefont {Zhang},
  \citenamefont {He}, \citenamefont {Wang}, \citenamefont {Lu}, \citenamefont
  {Ma},\ and\ \citenamefont {Xue}}]{science167}%
  \BibitemOpen
  \bibfield  {author} {\bibinfo {author} {\bibfnamefont {C.-Z.}\ \bibnamefont
  {Chang}}, \bibinfo {author} {\bibfnamefont {J.}~\bibnamefont {Zhang}},
  \bibinfo {author} {\bibfnamefont {X.}~\bibnamefont {Feng}}, \bibinfo {author}
  {\bibfnamefont {J.}~\bibnamefont {Shen}}, \bibinfo {author} {\bibfnamefont
  {Z.}~\bibnamefont {Zhang}}, \bibinfo {author} {\bibfnamefont
  {M.}~\bibnamefont {Guo}}, \bibinfo {author} {\bibfnamefont {K.}~\bibnamefont
  {Li}}, \bibinfo {author} {\bibfnamefont {Y.}~\bibnamefont {Ou}}, \bibinfo
  {author} {\bibfnamefont {P.}~\bibnamefont {Wei}}, \bibinfo {author}
  {\bibfnamefont {L.-L.}\ \bibnamefont {Wang}}, \bibinfo {author}
  {\bibfnamefont {Z.-Q.}\ \bibnamefont {Ji}}, \bibinfo {author} {\bibfnamefont
  {Y.}~\bibnamefont {Feng}}, \bibinfo {author} {\bibfnamefont {S.}~\bibnamefont
  {Ji}}, \bibinfo {author} {\bibfnamefont {X.}~\bibnamefont {Chen}}, \bibinfo
  {author} {\bibfnamefont {J.}~\bibnamefont {Jia}}, \bibinfo {author}
  {\bibfnamefont {X.}~\bibnamefont {Dai}}, \bibinfo {author} {\bibfnamefont
  {Z.}~\bibnamefont {Fang}}, \bibinfo {author} {\bibfnamefont {S.-C.}\
  \bibnamefont {Zhang}}, \bibinfo {author} {\bibfnamefont {K.}~\bibnamefont
  {He}}, \bibinfo {author} {\bibfnamefont {Y.}~\bibnamefont {Wang}}, \bibinfo
  {author} {\bibfnamefont {L.}~\bibnamefont {Lu}}, \bibinfo {author}
  {\bibfnamefont {X.-C.}\ \bibnamefont {Ma}}, \ and\ \bibinfo {author}
  {\bibfnamefont {Q.-K.}\ \bibnamefont {Xue}},\ }\href {\doibase
  10.1126/science.1234414} {\bibfield  {journal} {\bibinfo  {journal}
  {Science}\ }\textbf {\bibinfo {volume} {340}},\ \bibinfo {pages} {167}
  (\bibinfo {year} {2013})}\BibitemShut {NoStop}%
\bibitem [{\citenamefont {Yu}\ \emph {et~al.}(2010)\citenamefont {Yu},
  \citenamefont {Zhang}, \citenamefont {Zhang}, \citenamefont {Zhang},
  \citenamefont {Dai},\ and\ \citenamefont {Fang}}]{science61}%
  \BibitemOpen
  \bibfield  {author} {\bibinfo {author} {\bibfnamefont {R.}~\bibnamefont
  {Yu}}, \bibinfo {author} {\bibfnamefont {W.}~\bibnamefont {Zhang}}, \bibinfo
  {author} {\bibfnamefont {H.-J.}\ \bibnamefont {Zhang}}, \bibinfo {author}
  {\bibfnamefont {S.-C.}\ \bibnamefont {Zhang}}, \bibinfo {author}
  {\bibfnamefont {X.}~\bibnamefont {Dai}}, \ and\ \bibinfo {author}
  {\bibfnamefont {Z.}~\bibnamefont {Fang}},\ }\href {\doibase
  10.1126/science.1189926} {\bibfield  {journal} {\bibinfo  {journal}
  {Science}\ }\textbf {\bibinfo {volume} {329}},\ \bibinfo {pages} {61}
  (\bibinfo {year} {2010})}\BibitemShut {NoStop}%
\bibitem [{\citenamefont {Chen}\ \emph {et~al.}(2010)\citenamefont {Chen},
  \citenamefont {Chu}, \citenamefont {Analytis}, \citenamefont {Liu},
  \citenamefont {Igarashi}, \citenamefont {Kuo}, \citenamefont {Qi},
  \citenamefont {Mo}, \citenamefont {Moore}, \citenamefont {Lu}, \citenamefont
  {Mashimoto}, \citenamefont {Sasagawa}, \citenamefont {Zhang}, \citenamefont
  {Fisher}, \citenamefont {Hussain},\ and\ \citenamefont {Shen}}]{science659}%
  \BibitemOpen
  \bibfield  {author} {\bibinfo {author} {\bibfnamefont {Y.~L.}\ \bibnamefont
  {Chen}}, \bibinfo {author} {\bibfnamefont {J.-H.}\ \bibnamefont {Chu}},
  \bibinfo {author} {\bibfnamefont {J.~G.}\ \bibnamefont {Analytis}}, \bibinfo
  {author} {\bibfnamefont {Z.~K.}\ \bibnamefont {Liu}}, \bibinfo {author}
  {\bibfnamefont {K.}~\bibnamefont {Igarashi}}, \bibinfo {author}
  {\bibfnamefont {H.-H.}\ \bibnamefont {Kuo}}, \bibinfo {author} {\bibfnamefont
  {X.~L.}\ \bibnamefont {Qi}}, \bibinfo {author} {\bibfnamefont {S.~K.}\
  \bibnamefont {Mo}}, \bibinfo {author} {\bibfnamefont {R.~G.}\ \bibnamefont
  {Moore}}, \bibinfo {author} {\bibfnamefont {D.~H.}\ \bibnamefont {Lu}},
  \bibinfo {author} {\bibfnamefont {M.}~\bibnamefont {Mashimoto}}, \bibinfo
  {author} {\bibfnamefont {T.}~\bibnamefont {Sasagawa}}, \bibinfo {author}
  {\bibfnamefont {S.~C.}\ \bibnamefont {Zhang}}, \bibinfo {author}
  {\bibfnamefont {I.~R.}\ \bibnamefont {Fisher}}, \bibinfo {author}
  {\bibfnamefont {Z.}~\bibnamefont {Hussain}}, \ and\ \bibinfo {author}
  {\bibfnamefont {Z.~X.}\ \bibnamefont {Shen}},\ }\href {\doibase
  10.1126/science.1189924} {\bibfield  {journal} {\bibinfo  {journal}
  {Science}\ }\textbf {\bibinfo {volume} {329}},\ \bibinfo {pages} {659}
  (\bibinfo {year} {2010})}\BibitemShut {NoStop}%
\bibitem [{\citenamefont {Katmis}\ \emph {et~al.}(2016)\citenamefont {Katmis},
  \citenamefont {Lauter}, \citenamefont {Nogueira}, \citenamefont {Assaf},
  \citenamefont {Jamer}, \citenamefont {Wei}, \citenamefont {Satpati},
  \citenamefont {Freeland}, \citenamefont {Eremin}, \citenamefont {Heiman},
  \citenamefont {Jarillo-Herrero},\ and\ \citenamefont {Moodera}}]{nature513}%
  \BibitemOpen
  \bibfield  {author} {\bibinfo {author} {\bibfnamefont {F.}~\bibnamefont
  {Katmis}}, \bibinfo {author} {\bibfnamefont {V.}~\bibnamefont {Lauter}},
  \bibinfo {author} {\bibfnamefont {F.~S.}\ \bibnamefont {Nogueira}}, \bibinfo
  {author} {\bibfnamefont {B.~H.}\ \bibnamefont {Assaf}}, \bibinfo {author}
  {\bibfnamefont {M.~E.}\ \bibnamefont {Jamer}}, \bibinfo {author}
  {\bibfnamefont {P.}~\bibnamefont {Wei}}, \bibinfo {author} {\bibfnamefont
  {B.}~\bibnamefont {Satpati}}, \bibinfo {author} {\bibfnamefont {J.~W.}\
  \bibnamefont {Freeland}}, \bibinfo {author} {\bibfnamefont {I.}~\bibnamefont
  {Eremin}}, \bibinfo {author} {\bibfnamefont {D.}~\bibnamefont {Heiman}},
  \bibinfo {author} {\bibfnamefont {P.}~\bibnamefont {Jarillo-Herrero}}, \ and\
  \bibinfo {author} {\bibfnamefont {J.~S.}\ \bibnamefont {Moodera}},\ }\href
  {\doibase 10.1038/nature17635} {\bibfield  {journal} {\bibinfo  {journal}
  {Nature}\ }\textbf {\bibinfo {volume} {533}},\ \bibinfo {pages} {513}
  (\bibinfo {year} {2016})}\BibitemShut {NoStop}%
\bibitem [{\citenamefont {Luo}\ and\ \citenamefont {Qi}(2013)}]{prb085431}%
  \BibitemOpen
  \bibfield  {author} {\bibinfo {author} {\bibfnamefont {W.}~\bibnamefont
  {Luo}}\ and\ \bibinfo {author} {\bibfnamefont {X.-L.}\ \bibnamefont {Qi}},\
  }\href {\doibase 10.1103/PhysRevB.87.085431} {\bibfield  {journal} {\bibinfo
  {journal} {Phys. Rev. B}\ }\textbf {\bibinfo {volume} {87}},\ \bibinfo
  {pages} {085431} (\bibinfo {year} {2013})}\BibitemShut {NoStop}%
\bibitem [{\citenamefont {Li}\ \emph {et~al.}(2019{\natexlab{b}})\citenamefont
  {Li}, \citenamefont {Yu}, \citenamefont {Xu}, \citenamefont {Zhang},\ and\
  \citenamefont {Huang}}]{pb77}%
  \BibitemOpen
  \bibfield  {author} {\bibinfo {author} {\bibfnamefont {P.}~\bibnamefont
  {Li}}, \bibinfo {author} {\bibfnamefont {J.}~\bibnamefont {Yu}}, \bibinfo
  {author} {\bibfnamefont {J.}~\bibnamefont {Xu}}, \bibinfo {author}
  {\bibfnamefont {L.}~\bibnamefont {Zhang}}, \ and\ \bibinfo {author}
  {\bibfnamefont {K.}~\bibnamefont {Huang}},\ }\href {\doibase
  10.1016/j.physb.2019.08.009} {\bibfield  {journal} {\bibinfo  {journal}
  {Physcia B}\ }\textbf {\bibinfo {volume} {573}},\ \bibinfo {pages} {77}
  (\bibinfo {year} {2019}{\natexlab{b}})}\BibitemShut {NoStop}%
\bibitem [{\citenamefont {Peng}\ and\ \citenamefont {Xu}(2019)}]{prb195431}%
  \BibitemOpen
  \bibfield  {author} {\bibinfo {author} {\bibfnamefont {Y.}~\bibnamefont
  {Peng}}\ and\ \bibinfo {author} {\bibfnamefont {Y.}~\bibnamefont {Xu}},\
  }\href {\doibase 10.1103/PhysRevB.99.195431} {\bibfield  {journal} {\bibinfo
  {journal} {Phys. Rev. B}\ }\textbf {\bibinfo {volume} {99}},\ \bibinfo
  {pages} {195431} (\bibinfo {year} {2019})}\BibitemShut {NoStop}%
\bibitem [{\citenamefont {Fu}\ \emph {et~al.}(2020)\citenamefont {Fu},
  \citenamefont {Liu},\ and\ \citenamefont {Yan}}]{sa0948}%
  \BibitemOpen
  \bibfield  {author} {\bibinfo {author} {\bibfnamefont {H.}~\bibnamefont
  {Fu}}, \bibinfo {author} {\bibfnamefont {C.-X.}\ \bibnamefont {Liu}}, \ and\
  \bibinfo {author} {\bibfnamefont {B.}~\bibnamefont {Yan}},\ }\href {\doibase
  10.1126/sciadv.aaz0948} {\bibfield  {journal} {\bibinfo  {journal} {Sci.
  Adv.}\ }\textbf {\bibinfo {volume} {6}},\ \bibinfo {pages} {eaaz0948}
  (\bibinfo {year} {2020})}\BibitemShut {NoStop}%
\bibitem [{\citenamefont {Chowdhury}\ \emph {et~al.}(2019)\citenamefont
  {Chowdhury}, \citenamefont {Garrity},\ and\ \citenamefont
  {Tavazza}}]{npjcm33}%
  \BibitemOpen
  \bibfield  {author} {\bibinfo {author} {\bibfnamefont {S.}~\bibnamefont
  {Chowdhury}}, \bibinfo {author} {\bibfnamefont {K.~F.}\ \bibnamefont
  {Garrity}}, \ and\ \bibinfo {author} {\bibfnamefont {F.}~\bibnamefont
  {Tavazza}},\ }\href {\doibase 10.1038/s41524-019-0168-1} {\bibfield
  {journal} {\bibinfo  {journal} {npj Comput. Mater.}\ }\textbf {\bibinfo
  {volume} {5}},\ \bibinfo {pages} {33} (\bibinfo {year} {2019})}\BibitemShut
  {NoStop}%
\bibitem [{\citenamefont {Murakami}\ \emph {et~al.}(2019)\citenamefont
  {Murakami}, \citenamefont {Nambu}, \citenamefont {Koretsune}, \citenamefont
  {Xiangyu}, \citenamefont {Yamamoto}, \citenamefont {Brown},\ and\
  \citenamefont {Kageyama}}]{prb195103}%
  \BibitemOpen
  \bibfield  {author} {\bibinfo {author} {\bibfnamefont {T.}~\bibnamefont
  {Murakami}}, \bibinfo {author} {\bibfnamefont {Y.}~\bibnamefont {Nambu}},
  \bibinfo {author} {\bibfnamefont {T.}~\bibnamefont {Koretsune}}, \bibinfo
  {author} {\bibfnamefont {G.}~\bibnamefont {Xiangyu}}, \bibinfo {author}
  {\bibfnamefont {T.}~\bibnamefont {Yamamoto}}, \bibinfo {author}
  {\bibfnamefont {C.~M.}\ \bibnamefont {Brown}}, \ and\ \bibinfo {author}
  {\bibfnamefont {H.}~\bibnamefont {Kageyama}},\ }\href {\doibase
  10.1103/PhysRevB.100.195103} {\bibfield  {journal} {\bibinfo  {journal}
  {Phys. Rev. B}\ }\textbf {\bibinfo {volume} {100}},\ \bibinfo {pages}
  {195103} (\bibinfo {year} {2019})}\BibitemShut {NoStop}%
\bibitem [{\citenamefont {Zhou}\ \emph {et~al.}(2020)\citenamefont {Zhou},
  \citenamefont {Tan}, \citenamefont {Yan}, \citenamefont {Fang}, \citenamefont
  {Shi},\ and\ \citenamefont {Weng}}]{prb085114}%
  \BibitemOpen
  \bibfield  {author} {\bibinfo {author} {\bibfnamefont {L.}~\bibnamefont
  {Zhou}}, \bibinfo {author} {\bibfnamefont {Z.}~\bibnamefont {Tan}}, \bibinfo
  {author} {\bibfnamefont {D.}~\bibnamefont {Yan}}, \bibinfo {author}
  {\bibfnamefont {Z.}~\bibnamefont {Fang}}, \bibinfo {author} {\bibfnamefont
  {Y.}~\bibnamefont {Shi}}, \ and\ \bibinfo {author} {\bibfnamefont
  {H.}~\bibnamefont {Weng}},\ }\href {\doibase 10.1103/PhysRevB.102.085114}
  {\bibfield  {journal} {\bibinfo  {journal} {Phys. Rev. B}\ }\textbf {\bibinfo
  {volume} {102}},\ \bibinfo {pages} {085114} (\bibinfo {year}
  {2020})}\BibitemShut {NoStop}%
\bibitem [{\citenamefont {Kresse}\ and\ \citenamefont
  {Furthm\"{u}ller}(1996)}]{prb11169}%
  \BibitemOpen
  \bibfield  {author} {\bibinfo {author} {\bibfnamefont {G.}~\bibnamefont
  {Kresse}}\ and\ \bibinfo {author} {\bibnamefont {Furthm\"{u}ller}},\ }\href
  {\doibase 10.1103/PhysRevB.54.11169} {\bibfield  {journal} {\bibinfo
  {journal} {Phys. Rev. B}\ }\textbf {\bibinfo {volume} {54}},\ \bibinfo
  {pages} {11169} (\bibinfo {year} {1996})}\BibitemShut {NoStop}%
\bibitem [{\citenamefont {Perdew}\ \emph {et~al.}(1996)\citenamefont {Perdew},
  \citenamefont {Burke},\ and\ \citenamefont {Ernzerhof}}]{prl3865}%
  \BibitemOpen
  \bibfield  {author} {\bibinfo {author} {\bibfnamefont {J.~P.}\ \bibnamefont
  {Perdew}}, \bibinfo {author} {\bibfnamefont {K.}~\bibnamefont {Burke}}, \
  and\ \bibinfo {author} {\bibfnamefont {M.}~\bibnamefont {Ernzerhof}},\ }\href
  {\doibase 10.1103/PhysRevLett.77.3865} {\bibfield  {journal} {\bibinfo
  {journal} {Phys. Rev. Lett.}\ }\textbf {\bibinfo {volume} {77}},\ \bibinfo
  {pages} {3865} (\bibinfo {year} {1996})}\BibitemShut {NoStop}%
\bibitem [{\citenamefont {Klime\u{s}}\ \emph {et~al.}(2010)\citenamefont
  {Klime\u{s}}, \citenamefont {Bowler},\ and\ \citenamefont
  {Michaelides}}]{jpcm022201}%
  \BibitemOpen
  \bibfield  {author} {\bibinfo {author} {\bibfnamefont {J.}~\bibnamefont
  {Klime\u{s}}}, \bibinfo {author} {\bibfnamefont {D.~R.}\ \bibnamefont
  {Bowler}}, \ and\ \bibinfo {author} {\bibfnamefont {A.}~\bibnamefont
  {Michaelides}},\ }\href {\doibase 10.1088/0953-8984/22/2/022201} {\bibfield
  {journal} {\bibinfo  {journal} {J. Phys.: Condens. Matter}\ }\textbf
  {\bibinfo {volume} {22}},\ \bibinfo {pages} {022201} (\bibinfo {year}
  {2010})}\BibitemShut {NoStop}%
\bibitem [{\citenamefont {Heyd}\ \emph {et~al.}(2003)\citenamefont {Heyd},
  \citenamefont {Scuseria},\ and\ \citenamefont {Ernzerhof}}]{jcp8207}%
  \BibitemOpen
  \bibfield  {author} {\bibinfo {author} {\bibfnamefont {J.}~\bibnamefont
  {Heyd}}, \bibinfo {author} {\bibfnamefont {G.}~\bibnamefont {Scuseria}}, \
  and\ \bibinfo {author} {\bibfnamefont {M.}~\bibnamefont {Ernzerhof}},\ }\href
  {\doibase 10.1063/1.1564060} {\bibfield  {journal} {\bibinfo  {journal} {J.
  Chem. Phys.}\ }\textbf {\bibinfo {volume} {118}},\ \bibinfo {pages} {8207}
  (\bibinfo {year} {2003})}\BibitemShut {NoStop}%
\bibitem [{\citenamefont {Heyd}\ \emph {et~al.}(2006)\citenamefont {Heyd},
  \citenamefont {Scuseria},\ and\ \citenamefont {Ernzerhof}}]{jcp219906}%
  \BibitemOpen
  \bibfield  {author} {\bibinfo {author} {\bibfnamefont {J.}~\bibnamefont
  {Heyd}}, \bibinfo {author} {\bibfnamefont {G.}~\bibnamefont {Scuseria}}, \
  and\ \bibinfo {author} {\bibfnamefont {M.}~\bibnamefont {Ernzerhof}},\ }\href
  {\doibase 10.1063/1.2204597} {\bibfield  {journal} {\bibinfo  {journal} {J.
  Chem. Phys.}\ }\textbf {\bibinfo {volume} {124}},\ \bibinfo {pages} {219906}
  (\bibinfo {year} {2006})}\BibitemShut {NoStop}%
\bibitem [{\citenamefont {Wu}\ \emph {et~al.}(2018)\citenamefont {Wu},
  \citenamefont {Zhang}, \citenamefont {Song}, \citenamefont {Troyer},\ and\
  \citenamefont {Soluyanov}}]{cpc405}%
  \BibitemOpen
  \bibfield  {author} {\bibinfo {author} {\bibfnamefont {Q.}~\bibnamefont
  {Wu}}, \bibinfo {author} {\bibfnamefont {S.}~\bibnamefont {Zhang}}, \bibinfo
  {author} {\bibfnamefont {H.-F.}\ \bibnamefont {Song}}, \bibinfo {author}
  {\bibfnamefont {M.}~\bibnamefont {Troyer}}, \ and\ \bibinfo {author}
  {\bibfnamefont {A.~A.}\ \bibnamefont {Soluyanov}},\ }\href {\doibase
  10.1016/j.cpc.2017.09.033} {\bibfield  {journal} {\bibinfo  {journal}
  {Comput. Phys. Commun.}\ }\textbf {\bibinfo {volume} {224}},\ \bibinfo
  {pages} {405} (\bibinfo {year} {2018})}\BibitemShut {NoStop}%
\bibitem [{\citenamefont {Nowka}\ \emph {et~al.}(2017)\citenamefont {Nowka},
  \citenamefont {Gellesch}, \citenamefont {Borrero}, \citenamefont {Partzsch},
  \citenamefont {Wuttke}, \citenamefont {Steckel}, \citenamefont {Hess},
  \citenamefont {Wolter}, \citenamefont {Bohorquez}, \citenamefont
  {B\"{u}chner},\ and\ \citenamefont {Hampel}}]{jcg81}%
  \BibitemOpen
  \bibfield  {author} {\bibinfo {author} {\bibfnamefont {C.}~\bibnamefont
  {Nowka}}, \bibinfo {author} {\bibfnamefont {M.}~\bibnamefont {Gellesch}},
  \bibinfo {author} {\bibfnamefont {J.~E.~H.}\ \bibnamefont {Borrero}},
  \bibinfo {author} {\bibfnamefont {S.}~\bibnamefont {Partzsch}}, \bibinfo
  {author} {\bibfnamefont {C.}~\bibnamefont {Wuttke}}, \bibinfo {author}
  {\bibfnamefont {F.}~\bibnamefont {Steckel}}, \bibinfo {author} {\bibfnamefont
  {C.}~\bibnamefont {Hess}}, \bibinfo {author} {\bibfnamefont {A.~U.~B.}\
  \bibnamefont {Wolter}}, \bibinfo {author} {\bibfnamefont {L.~T.~C.}\
  \bibnamefont {Bohorquez}}, \bibinfo {author} {\bibfnamefont {B.}~\bibnamefont
  {B\"{u}chner}}, \ and\ \bibinfo {author} {\bibfnamefont {S.}~\bibnamefont
  {Hampel}},\ }\href {\doibase 10.1016/j.jcrysgro.2016.11.090} {\bibfield
  {journal} {\bibinfo  {journal} {J. Cryst. Growth}\ }\textbf {\bibinfo
  {volume} {459}},\ \bibinfo {pages} {81} (\bibinfo {year} {2017})}\BibitemShut
  {NoStop}%
\bibitem [{\citenamefont {Li}\ \emph {et~al.}(2019{\natexlab{c}})\citenamefont
  {Li}, \citenamefont {Yu}, \citenamefont {Xu}, \citenamefont {Zhang},\ and\
  \citenamefont {Huang}}]{aqt201900033}%
  \BibitemOpen
  \bibfield  {author} {\bibinfo {author} {\bibfnamefont {P.}~\bibnamefont
  {Li}}, \bibinfo {author} {\bibfnamefont {J.}~\bibnamefont {Yu}}, \bibinfo
  {author} {\bibfnamefont {J.}~\bibnamefont {Xu}}, \bibinfo {author}
  {\bibfnamefont {L.}~\bibnamefont {Zhang}}, \ and\ \bibinfo {author}
  {\bibfnamefont {K.}~\bibnamefont {Huang}},\ }\href {\doibase
  10.1002/qute.201900033} {\bibfield  {journal} {\bibinfo  {journal} {Adv.
  Quantum Technol.}\ }\textbf {\bibinfo {volume} {2}},\ \bibinfo {pages}
  {201900033} (\bibinfo {year} {2019}{\natexlab{c}})}\BibitemShut {NoStop}%
\bibitem [{\citenamefont {Li}\ and\ \citenamefont {Luo}(2016)}]{sr25423}%
  \BibitemOpen
  \bibfield  {author} {\bibinfo {author} {\bibfnamefont {P.}~\bibnamefont
  {Li}}\ and\ \bibinfo {author} {\bibfnamefont {W.}~\bibnamefont {Luo}},\
  }\href {\doibase 10.1038/srep25423} {\bibfield  {journal} {\bibinfo
  {journal} {Sci. Rep.}\ }\textbf {\bibinfo {volume} {6}},\ \bibinfo {pages}
  {25423} (\bibinfo {year} {2016})}\BibitemShut {NoStop}%
\bibitem [{\citenamefont {Hasan}\ and\ \citenamefont {Kane}(2010)}]{rmp3045}%
  \BibitemOpen
  \bibfield  {author} {\bibinfo {author} {\bibfnamefont {M.~Z.}\ \bibnamefont
  {Hasan}}\ and\ \bibinfo {author} {\bibfnamefont {C.~L.}\ \bibnamefont
  {Kane}},\ }\href {\doibase 10.1103/RevModPhys.82.3045} {\bibfield  {journal}
  {\bibinfo  {journal} {Rev. Mod. Phys.}\ }\textbf {\bibinfo {volume} {82}},\
  \bibinfo {pages} {3045} (\bibinfo {year} {2010})}\BibitemShut {NoStop}%
\bibitem [{\citenamefont {Qi}\ and\ \citenamefont {Zhang}(2011)}]{rmp1057}%
  \BibitemOpen
  \bibfield  {author} {\bibinfo {author} {\bibfnamefont {X.-L.}\ \bibnamefont
  {Qi}}\ and\ \bibinfo {author} {\bibfnamefont {S.-C.}\ \bibnamefont {Zhang}},\
  }\href {\doibase 10.1103/RevModPhys.83.1057} {\bibfield  {journal} {\bibinfo
  {journal} {Rev. Mod. Phys.}\ }\textbf {\bibinfo {volume} {83}},\ \bibinfo
  {pages} {1057} (\bibinfo {year} {2011})}\BibitemShut {NoStop}%
\bibitem [{\citenamefont {Shi}\ \emph {et~al.}(2020)\citenamefont {Shi},
  \citenamefont {Zhang}, \citenamefont {Yan}, \citenamefont {Feng},
  \citenamefont {Yang}, \citenamefont {Shi},\ and\ \citenamefont
  {Li}}]{cpl047301}%
  \BibitemOpen
  \bibfield  {author} {\bibinfo {author} {\bibfnamefont {G.}~\bibnamefont
  {Shi}}, \bibinfo {author} {\bibfnamefont {M.}~\bibnamefont {Zhang}}, \bibinfo
  {author} {\bibfnamefont {D.}~\bibnamefont {Yan}}, \bibinfo {author}
  {\bibfnamefont {H.}~\bibnamefont {Feng}}, \bibinfo {author} {\bibfnamefont
  {M.}~\bibnamefont {Yang}}, \bibinfo {author} {\bibfnamefont {Y.}~\bibnamefont
  {Shi}}, \ and\ \bibinfo {author} {\bibfnamefont {Y.}~\bibnamefont {Li}},\
  }\href {\doibase 10.1088/0256-307X/37/4/047301} {\bibfield  {journal}
  {\bibinfo  {journal} {Chin. Phys. Lett.}\ }\textbf {\bibinfo {volume} {37}},\
  \bibinfo {pages} {047301} (\bibinfo {year} {2020})}\BibitemShut {NoStop}%
\bibitem [{\citenamefont {\v{S}mejkal}\ \emph {et~al.}(2018)\citenamefont
  {\v{S}mejkal}, \citenamefont {Mokrousov}, \citenamefont {Yan},\ and\
  \citenamefont {MacDonald}}]{np242}%
  \BibitemOpen
  \bibfield  {author} {\bibinfo {author} {\bibfnamefont {L.}~\bibnamefont
  {\v{S}mejkal}}, \bibinfo {author} {\bibfnamefont {Y.}~\bibnamefont
  {Mokrousov}}, \bibinfo {author} {\bibfnamefont {B.}~\bibnamefont {Yan}}, \
  and\ \bibinfo {author} {\bibfnamefont {A.~H.}\ \bibnamefont {MacDonald}},\
  }\href {\doibase 10.1038/s41567-018-0064-5} {\bibfield  {journal} {\bibinfo
  {journal} {Nat. Phys.}\ }\textbf {\bibinfo {volume} {14}},\ \bibinfo {pages}
  {242} (\bibinfo {year} {2018})}\BibitemShut {NoStop}%
\bibitem [{\citenamefont {Ge}\ \emph {et~al.}(2020)\citenamefont {Ge},
  \citenamefont {Liu}, \citenamefont {Li}, \citenamefont {Li}, \citenamefont
  {Luo}, \citenamefont {Wu}, \citenamefont {Xu},\ and\ \citenamefont
  {Wang}}]{nsr1280}%
  \BibitemOpen
  \bibfield  {author} {\bibinfo {author} {\bibfnamefont {J.}~\bibnamefont
  {Ge}}, \bibinfo {author} {\bibfnamefont {Y.}~\bibnamefont {Liu}}, \bibinfo
  {author} {\bibfnamefont {J.}~\bibnamefont {Li}}, \bibinfo {author}
  {\bibfnamefont {H.}~\bibnamefont {Li}}, \bibinfo {author} {\bibfnamefont
  {T.}~\bibnamefont {Luo}}, \bibinfo {author} {\bibfnamefont {Y.}~\bibnamefont
  {Wu}}, \bibinfo {author} {\bibfnamefont {Y.}~\bibnamefont {Xu}}, \ and\
  \bibinfo {author} {\bibfnamefont {J.}~\bibnamefont {Wang}},\ }\href {\doibase
  10.1093/nsr/nwaa089} {\bibfield  {journal} {\bibinfo  {journal} {Natl. Sci.
  Rev.}\ }\textbf {\bibinfo {volume} {7}},\ \bibinfo {pages} {1280} (\bibinfo
  {year} {2020})}\BibitemShut {NoStop}%
\end{thebibliography}%
\end{document}